\documentclass[preprint,3p]{elsarticle}

\usepackage{graphicx}
\usepackage{caption}
\usepackage{subcaption}
\usepackage{comment}
\usepackage{amssymb}
\usepackage{amsmath}
\usepackage{dirtytalk}
\usepackage{algorithm}
\usepackage{algorithmic}
\biboptions{numbers,sort&compress}
\usepackage{lineno}

\usepackage{xcolor}
\usepackage{hyperref}
\usepackage{siunitx}
\usepackage{tikz}
\usepackage{pgfplots}
\usepackage{color,soul}
\usetikzlibrary{matrix}
\usetikzlibrary{plotmarks}
\pgfplotsset{compat=newest}
\usepackage{makecell}
\usepackage[export]{adjustbox}

\newcommand{\mc}[1]{\mathcal{#1}}

\newcommand{\bs}[1]{\boldsymbol{#1}}

\usepackage{amsmath}

\DeclareMathOperator*{\argmin}{arg\,min}

\hypersetup{      
    colorlinks=true,                
    linkcolor= blue,                
    filecolor=black,
    citecolor=blue,
    linkbordercolor=blue
}
\graphicspath{ {./Figures/} }
\definecolor{brandeisblue}{rgb}{0.0, 0.44, 1.0}
\definecolor{brightpink}{rgb}{1.0, 0.0, 0.5}
\journal{Elsevier}
\begin{document}

\setlength{\baselineskip}{14pt}

\begin{frontmatter}


\title{Application of Physics-Informed Neural Networks for Forward and Inverse Analysis of Pile-Soil Interaction}



\author[civilUNSW]{M. Vahab}
\author[civilUNSW]{B. Shahbodagh}
\cortext[cor]{Corresponding author:}
\ead{b.shahbodagh@unsw.edu.au}
\author[mechanicalMIT]{E. Haghighat}
\author[civilUNSW]{N. Khalili}

\address[civilUNSW]{School of Civil and Environmental Engineering, The University of New South Wales, Sydney 2052, Australia}
\address[mechanicalMIT]{Massachusetts Institute of Technology, Cambridge, MA, USA}

\begin{abstract}
The application of the Physics-Informed Neural Networks (PINNs) to forward and inverse analysis of pile-soil interaction problems is presented. The main challenge encountered in the Artificial Neural Network (ANN) modelling of pile-soil interaction is the presence of abrupt changes in material properties, which results in large discontinuities in the gradient of the displacement solution. Therefore, a domain-decomposition multi-network model is proposed to deal with the discontinuities in the strain fields at common boundaries of pile-soil regions and soil layers.
The application of the model to the analysis and parametric study of single piles embedded in both homogeneous and layered formations is demonstrated under axisymmetric and plane strain conditions. The performance of the model in parameter identification (inverse analysis) of pile-soil interaction is particularly investigated. It is shown that by using PINNs, the localized data acquired along the pile length \textendash possibly obtained via fiber optic strain sensing\textendash can be successfully used for the inversion of soil parameters in layered formations.

\end{abstract}

\begin{keyword}
Physics-Informed Neural Networks (PINNs); Deep learning; Pile-soil interaction; SciANN



\end{keyword}

\end{frontmatter}

\section{Introduction}
\label{S:1 (introduction)}
\subsection{Background}

The Finite Element Method (FEM) is ubiquitously used for the computational analysis of various geotechnical engineering problems \cite{zienkiewicz1999computational, ghasemi2013practical, rahmani2012dynamic,jafari2021fully}. Consistently, in many classical studies, FEM has been the natural choice for the design of inversion or back-analysis algorithms for inferring important model parameters, like mechanical properties of soils \cite{calvello2004selecting}.
The state-of-the-art techniques in this context mainly rely on minimization of the deviations between the simulation responses (due to an adjustable set of model parameters) with respect to field and laboratory measurements. For this purpose, a range of direct or gradient-based optimization techniques are extended which exploit successive numerical solutions iteratively to yield optimal precision for the sought-after parameters \cite{kabe1985stiffness}. Nonetheless, both approaches can be prohibitively computationally expensive in particular for large-scale problems. 
Further challenges encountered during numerical back analysis in engineering applications have been due to non-uniqueness, material model limitations, and disparate data sources \cite{vardakos2012parameter, walton2022challenges}.

Deep Learning (DL) has proven to be a rigorous approach for the forward and back analysis of geotechnical engineering problems \cite{lefik2002artificial, gawin2001ann, parish2016paradigm, kardani2020estimation}. In the analysis of piles, which is the focus of this study, DL has been applied for the prediction of the shaft and tip resistance of concrete piles by Momeni et al. \cite{momeni2015application}, for estimation of the uplift resistance of screw piles by Mosallanezhad and Moayedi \cite{moayedi2017uplift}, and for evaluation of the lateral load bearing capacity of piles by Das and Basudhar \cite{das2006undrained} and Armaghan et al. \cite{armaghani2017developing}. It is noteworthy that the enumerated studies emphasize on forward prediction of mechanical response of piles itself. Based on the Universal approximation theorem (a.k.a., Cybenko theorem \cite{cybenko1989approximation}), neural networks comprising of at least a single hidden layer can uniformly
reconstruct any function with arbitrary continuous non-linearity. Evidently, the composite system of soil-piles involves a strong material discontinuity, which in turn leads to a discontinuous strain field (i.e., the gradient of displacement). 
This study aims to develop a DL framework capable of handling the discontinuities induced by material interfaces within soil-pile systems.

\subsection{Physics-Informed Deep Learning}
Early Artificial Intelligence (AI) approaches were developed in the mid-$20^\text{th}$ century to tackle the intellectually difficult and complex-to-articulate problems, which pertain to straight-forward solution algorithms within relatively sterile environments \cite{bishop2006pattern}. 
DL aims to mitigate major sources of difficulty with conventional real-world Machine Learning (ML) practice, in terms of data sensitivity, noises, and other representation learning issues, through a nested hierarchy of simpler representations/concepts \cite{goodfellow2016deep}. 
 Nowadays DL algorithms are predominantly employed in an increasing number of areas in engineering and science, including geotechnics \cite{zhang2021application}, structural engineering \cite{azimi2020structural,bao2019computer}, engineering mechanics \cite{li2019predicting}, reservoir engineering \cite{zhang2018modeling}, material science \cite{azimi2018advanced}, physical modelling \cite{de2019deep}, and earth sciences \cite{reichstein2019deep}, to name a few.

Most conventional DL algorithms rely on the availability of data for training to enable the development of reliable predictive models. However, in many engineering circumstances, the data acquisition costs could be prohibitive. 
Partial information threatens the robustness of machine learning to the extent that draws decision-making cumbersome, if not impossible. 
In a diversity of engineering and scientific applications, there exists a vast prior physical knowledge that can be exercised as a regularization agent rendering the admissible solution space into a manageable size \cite{ karniadakis2021physics}. These approaches could be considered as a subset of Reinforcement Learning (RL) \cite{sutton2018reinforcement} in a sense that a range of incentives are optimized rather than relying on the training dataset alone \cite{han2018solving,haghighat2021physics}. Concurrent efforts in recent years have been made to incorporate prior physical information in DL by Owhadi \cite{owhadi2015bayesian}, Han et al. \cite{han2018solving}, Bar-Sinai et al. \cite{bar2019learning}, Rudy et al. \cite{rudy2017data}, and Raissi et al. \cite{raissi2019physics}.

Physics-Informed Neural Networks (PINNs) are a class of deep learning that incorporate a series of physical laws, frequently described in the form of partial differential equations (PDEs), to steer the learning towards the solution for sparse training data-sets, which could not be plausible with classic DL algorithms. The prosperity of PINNs is attributed to substantial algorithmic advances (e.g., graph-based automated differentiation \cite{baydin2018automatic}) and major software developments (e.g., TensorFlow \cite{abadi2016tensorflow}, Keras \cite{chollet2015keras}). 
PINNs have been utilized with great success in a broad range of engineering disciplines, including solid mechanics \cite{haghighat2021sciann, haghighat2021physics, haghighat2021nonlocal, vahab2022physics, Vahab2022}, fluid mechanics \cite{mao2020physics, sahli2020physics, kharazmi2021hp}, and thermo-mechanics \cite{niaki2021physics}.
Conventional data-driven inversion models frequently fail in generalization, owing to extrapolation or observational biases \cite{ karniadakis2021physics,raissi2019physics}. 
A beneficial remedy can be offered by physics-informed deep learning, which provides the network model along with the laws of physics governing the system. This can rectify the issues associated with the missing ingredients induced by the sparsity of data, uncertainties, or other less understood factors.
 Inverse physics-informed solutions have been hitherto investigated in nano optics by Chen et al. \cite{chen2020physics}, in solid mechanics by Haghighat et al. \cite{haghighat2021physics}, for conservation laws by Jagtap et al. \cite{jagtap2020conservative}, and for flow problems by Lou et al. \cite{lou2021physics}.

\subsection{Our contribution}

In this study, we emphasize on a novel application of PINNs to the solution and inverse analysis of soil-pile interaction problems. For this sake, we use the longitudinal strain profile along the entire length of piles, which could be obtained through the optical fiber strain-sensing technique in practice (e.g., see Mohamad et al. \cite{mohamad2011performance,mohamad2012monitoring}). 
Differential strains (i.e., at either side of piles) may also be extracted by the installation of a group of fibers so as to enable the monitoring of both axial movements and bending \cite{mohamad2011performance}. 
Here, PINNs are first employed to construct efficient neural networks for the solution of soil-pile interactions in the absence of any data. Salient features of these systems, namely strain discontinuity, contact constraints, compatibility of stresses, and inhomogeneities, are incorporated through introducing a domain-decomposition multi-network model \cite{kharazmi2021hp, niaki2021physics, jagtap2020conservative, jagtap2021extended}. The localized data acquired along the pile length, in turn, is utilized for the inversion of key mechanical properties of soil in layered formations.

This paper is organized as follows:  In section \ref{S:2 (GoverningEq)}, the equations governing the pile-soil interactions are explained in detail. Section \ref{S:3 (PINN)} is devoted to a brief introduction to the fundamentals of PINNs and its application to the solution of pile-soil systems. The forward solution of single piles \textendash in both cylindrical and Cartesian coordinate systems \textendash is explored in section \ref{S:4 (solutions)}. In addition, the inverse analysis of soil mechanical properties is conducted for homogeneous and layered formations. Concluding remarks are presented in section \ref{S:5 (Conclusions)}.

\section{Problem Statement and Governing Equations}
\label{S:2 (GoverningEq)}

\begin{figure}[!b]
\centering
\begin{minipage}{0.75\textwidth}
    \centering
    \includegraphics[width=0.8\linewidth]{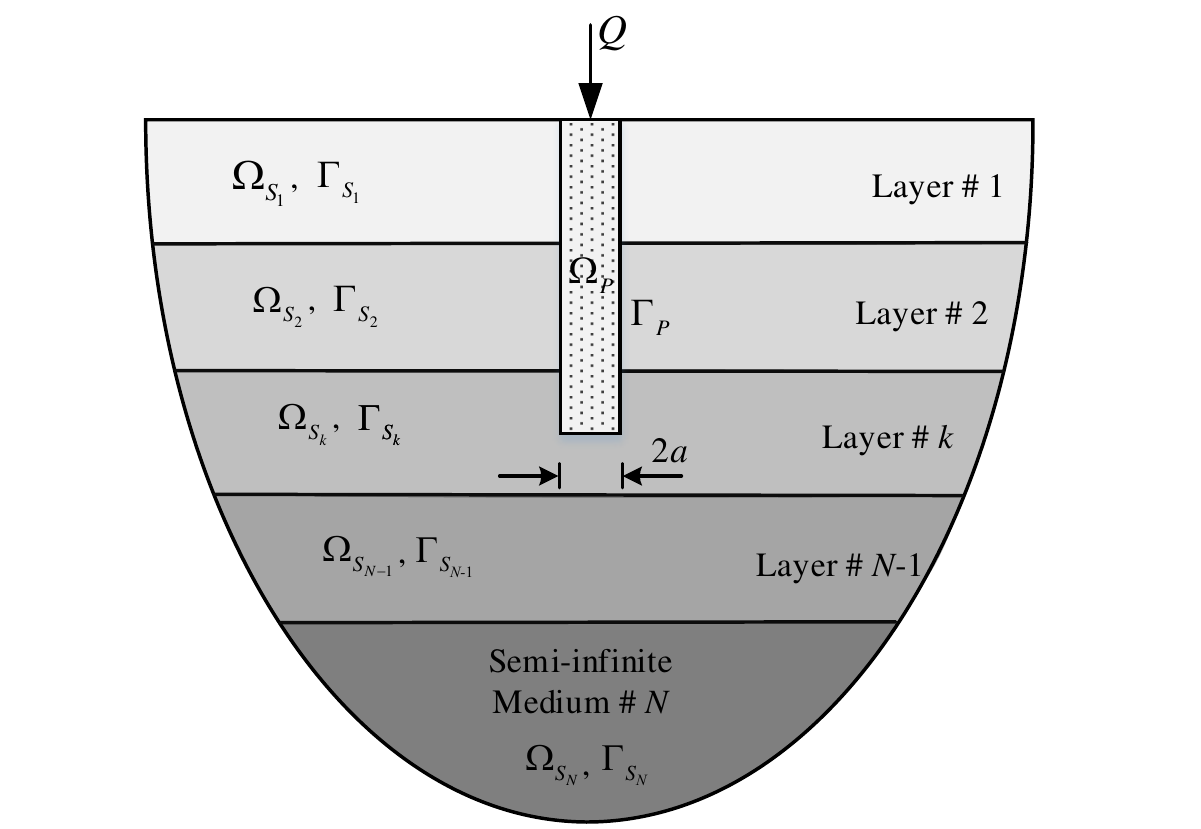}
    \end{minipage}
    \caption{Schematic representation of a single pile in a layered soil medium.}
    \label{fig:layered}
\end{figure} 

Fig. \ref{fig:layered} depicts the pile-soil system considered in this study. The system is modelled in both two-dimensional (2D) and three-dimensional (3D) settings, representing a sheet-pile wall and a single cylindrical pile embedded in soil media, respectively. 
The Cartesian coordinate system is used for the 2D plane strain analysis, whereas the cylindrical coordinate system is employed for the 3D analysis of pile-soil interaction. In the figure, ${{\Omega }_{P}}$ and ${{\Omega }_{S_k}} (k=1,..., N)$ denote the regions of the space occupied by the pile and the $k^\text{th}$ soil layer, respectively, where \textit{N} is the number of the soil layers. 
The boundaries of the pile and soil regions are designated by $\Gamma_p $ and $\Gamma_{S_k}$, respectively. The pile is assumed to be continuously bonded to and fully embedded in the soil medium. 
We leverage a generic representation of the pile deformation which enables the model to capture the non-uniform deformation along the cross-section of the pile, essential for the proper simulation of short pile response and interactions at low pile-soil stiffness ratios. The equilibrium equations governing the pile-soil system are expressed in the cylindrical coordinate system as
\begin{equation}
\begin{gathered}
\begin{aligned}
\label{eq:Equilibrium}
  & \frac{1}{r}\frac{\partial }{\partial r}\left( r{{\sigma }^\alpha_{rr}} \right)+\frac{1}{r}\frac{\partial {{\sigma }^\alpha_{r\theta }}}{\partial \theta }+\frac{\partial {{\sigma }^\alpha_{rz}}}{\partial z}-\frac{{{\sigma }^\alpha_{\theta \theta }}}{r}+{{f}^\alpha_{r}}=0\,\,\,, \\ 
 &  \\ 
 & \frac{1}{r}\frac{\partial }{\partial r}\left( r{{\sigma }^\alpha_{r\theta }} \right)+\frac{1}{r}\frac{\partial {{\sigma }^\alpha_{\theta \theta }}}{\partial \theta }+\frac{\partial {{\sigma }^\alpha_{\theta z}}}{\partial z}+\frac{{{\sigma }^\alpha_{r\theta }}}{r}+{{f}^\alpha_{\theta }}=0\,\,\,,\,\,\,\,\,\,\,\,\,\,\,\,\,\,\text{on}\,\,{{\Omega }_{\alpha }}\,\left( \alpha =P,\ S_k \right) \\ 
 &  \\ 
 & \frac{1}{r}\frac{\partial }{\partial r}\left( r{{\sigma }^\alpha_{rz}} \right)+\frac{1}{r}\frac{\partial {{\sigma }^\alpha_{\theta z}}}{\partial \theta }+\frac{\partial {{\sigma }^\alpha_{zz}}}{\partial z}+{{f}^\alpha_{z}}=0\,\,\,, \\ 
\end{aligned}
\end{gathered}
\end{equation}
where $\sigma^\alpha_{ij} \,(i,j=r,\theta ,z)$ is the Cauchy stress tensor , $f^\alpha_{i} $ is the body-force density vector, and \textit{P} and $S_k (k=1,..., N)$ denote the corresponding quantities of the pile region and the soil medium, respectively. The strain-displacement relations are given by

\begin{equation}
\begin{gathered}
\begin{aligned}
\label{eq:StrainDisp}
  & {{\varepsilon }^\alpha_{rr}}=\frac{\partial {{u}^\alpha_{r}}}{\partial r}\,\,\,,\,\,\,\,\,\,\,{{\varepsilon }^\alpha_{\theta \theta }}=\frac{1}{r}\left( \frac{\partial {{u}^\alpha_{\theta }}}{\partial \theta }+{{u}^\alpha_{r}} \right)\,\,\,,\,\,\,\,\,\,\,{{\varepsilon }^\alpha_{zz}}=\frac{\partial {{u}^\alpha_{z}}}{\partial z}\,\,\,, \\ 
 &  \\ 
 & {{\varepsilon }^\alpha_{r\theta }}=\frac{1}{2}\left( \frac{1}{r}\frac{\partial {{u}^\alpha_{r}}}{\partial \theta }+r\frac{\partial }{\partial r}\left( \frac{{{u}^\alpha_{\theta }}}{r} \right) \right)\,\,\,,\,\,\,\,\,\,\,{{\varepsilon }^\alpha_{z\theta }}=\frac{1}{2}\left( \frac{1}{r}\frac{\partial {{u}^\alpha_{z}}}{\partial \theta }+\frac{\partial {{u}^\alpha_{\theta }}}{\partial z} \right)\,\,\,,\,\,\,\,\,\,\,\,\text{on}\,\,{{\Omega }_{\alpha }}\,\left( \alpha =P,\ S_k \right) \\ 
 &  \\ 
 & {{\varepsilon }^\alpha_{rz}}=\frac{1}{2}\left( \frac{\partial {{u}^\alpha_{r}}}{\partial z}+\frac{\partial {{u}^\alpha_{z}}}{\partial r} \right)\,\,\,, \\ 
\end{aligned}
\end{gathered}
\end{equation}
where $\varepsilon^\alpha_{{ij}}\,\,(i,j=r,\theta ,z)$ is the strain tensor and $u^\alpha_{i} $ is the displacement vector. Using the indicial notation, Eq.s \eqref{eq:Equilibrium} and \eqref{eq:StrainDisp} can be written in the Cartesian coordinate system $(i,j=x, y, z)$ as
\begin{equation}
\begin{gathered}
\begin{aligned}
\label{GovEqCar}
\begin{aligned}
& \sigma^\alpha _{ji,j}+{{f}^\alpha_{i}}=0\,\,\,, \\
 & \varepsilon^\alpha _{ij}=\frac{1}{2}\left( u^\alpha_{i,j}+u^\alpha_{j,i} \right)\,\,\,, 
\end{aligned}
\,\,\,\,\,\,\,\,\,\,\,\,\,\,\,\,\,\,\,\,\,\,\,\text{on}\,\,{{\Omega }_{\alpha }}\,\,\left( \alpha =P,\ S_k \right) 
\end{aligned}
\end{gathered}
\end{equation}

The elastic constitutive relation employed for the pile-soil system is given by
\begin{equation}
\begin{gathered}
\begin{aligned}
\label{eq:Constitutive}
\sigma^\alpha _{ij} ={{\lambda }_{\alpha }}{{\varepsilon }^\alpha_{kk}}{{\delta }_{ij}}+2{{\mu }_{\alpha }}{{\varepsilon }^\alpha_{ij}}\,\,\,,\,\,\,\,\,\,\,\,\,\,\,\,\,\text{on}\,\,{{\Omega }_{\alpha }}\,\left( \alpha =P,\ S_k \right)
\end{aligned}
\end{gathered}
\end{equation}
where ${{\lambda }_{\alpha }}$ and ${{\mu }_{\alpha }}$ are the Lamé constants and ${{\delta }_{ij}}$ is the Kronecker delta. The Lamé constants can be expressed in terms of the Young's Modulus $E_{\alpha}$ and Poisson's ratio $\nu_{\alpha}$ as:
\begin{equation}
\begin{gathered}
\begin{aligned}
\label{eq:E&Nu}
\lambda_{\alpha} =\frac{E_{\alpha}\nu_{\alpha}}{(1+\nu_{\alpha})(1-2\nu_{\alpha})}\,\,\,, \ \ \ 
\mu_{\alpha} =\frac{E_{\alpha}}{2(1+\nu_{\alpha})}\,\,\,.
\end{aligned}
\end{gathered}
\end{equation}

These field equations are accompanied by the boundary conditions at the interfaces of pile and soil and soil layers, i.e.




\begin{equation}
\begin{gathered}
\begin{aligned}
\label{BC10}
\begin{aligned}
\left(\sigma^\alpha_{ji} (\bs{x})-\sigma ^\beta_{ji} (\bs{x}) \right){{n}^{\alpha}_{j}}=0\,\,\,, \\
                  u^\alpha_i (\bs{x}) - u^\beta_i (\bs{x})=0\,\,\,, \\ 
\end{aligned}
\,\,\,\,\,\,\,\,\,\,\,\,\,\,\,\,\,\,\,\,\,\,\,\forall  \bs{x}\in (\Gamma_\alpha\cap\Gamma_\beta)  \text{, where} \,\,\ \alpha \neq \beta
\\ 
\end{aligned}
\end{gathered}
\end{equation}
in which ${{n}^{{\alpha}}_{j}}$ is the unit outward normal vector of $\Gamma_\alpha $ $\left( \alpha , \beta =P,\ S_k \right)$. On the top surface, 
\begin{equation}
\begin{gathered}
\begin{aligned}
\label{eq:BC2}
  & \sigma^{S_1} _{ji} \left( r>a ,\theta,z=0 \right){{n}^{S_{1}}_{j}}=0\,\,\,,\, \\ 
  & \sigma^P _{ji} \left( 0\le r\le a ,\theta, z=0 \right){{n}^{P}_{j}}={{t}^P_{i}}\,\,\,, \\ 
\end{aligned}
\end{gathered}
\end{equation}
where ${{t}^P_{i}}$ is the surface traction applied to the pile head and $a$ is the radius of the pile. Under vertical loading condition, ${{t}^P_{i}}=\left\{ \begin{matrix}
   Q/\left( \pi {{a}^{2}} \right) & 0 & 0  \\
\end{matrix} \right\}\,$, where $Q$ is the applied vertical load. 
For a single pile embedded in a half-space, the regularity conditions at infinity are specified as
\begin{equation}
\label{BC3}
\sigma^{S_k} _{ij} \left( r,\theta ,z \right)\to 0\,\,\,,\,\,\,\,\,\,\,\,\text{as}\,\,\,\sqrt{{{r}^{2}}+{{z}^{2}}}\to \infty 
\end{equation}
For the case with underlying bedrock, the boundary condition at the bedrock level can be expressed as 
\begin{equation}
\label{BC33}
u^{S_{N}}_{i} \left( r,\,\theta ,z=H \right)=0\,\,\,,
\end{equation}
where \textit{H} is the depth to bedrock.



\section{PINN Forward and Inverse Solvers}
\label{S:3 (PINN)}

In this section, we briefly review the construction and training process of Physics-Informed Neural Network (PINN) solvers. Next, we implement the PINNs for the analysis of piles in homogeneous and layered formations.

\subsection{Physics-Informed Neural Networks}
\label{S:3-1 (basics)}

\begin{figure}[!t]
\centering
\begin{minipage}{0.65\textwidth}
    \centering
    \includegraphics[width=0.8\linewidth]{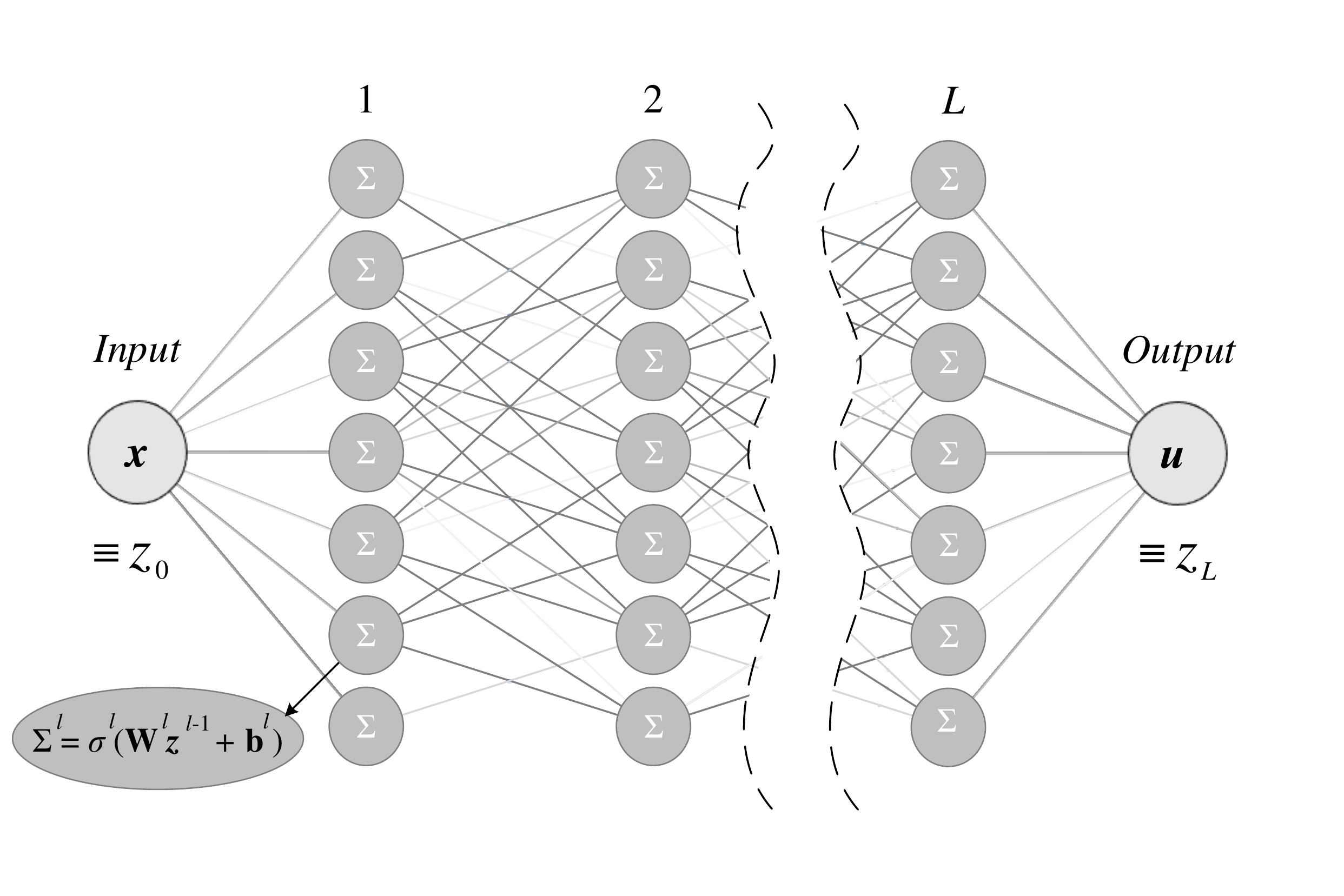}
    \end{minipage}
\caption{Standard PINN architecture, defining the mapping $\boldsymbol{u}:\boldsymbol{x}\mapsto\mathcal{N}_u(\boldsymbol{x};\mathbf{W},\mathbf{b})$.}
    \label{fig:network}
\end{figure} 

Suppose $\mathcal{N}_u(\boldsymbol{x};\mathbf{W},\mathbf{b})$ is an $L$-layer neural network with $\boldsymbol{x}$ and $\boldsymbol{u}$ being the input and output vectors, and $\mathbf{W}$ and $\mathbf{b}$ being weights and biases of the transformation, respectively. 
As depicted in  Fig. \ref{fig:network}, a feed-forward network is employed to approximate the solution variable $\boldsymbol{u}$ for given inputs $\boldsymbol{u}$ through the following transformations:
\begin{equation}
\label{eq:NN1}
\bs{u} = \Sigma^L \circ \Sigma^{L-1} \circ \dots \circ \Sigma^1(\bs{x})\,\,\,,
\end{equation}
where
\begin{equation}
\label{eq:ANN}
\Sigma^l(\hat{\bs{z}}^{l-1}) := \hat{\bs{z}}^l = \sigma^l(\mathbf{W}^l \cdot \hat{\bs{z}}^{l-1} + \mathbf{b}^l)\,\,\,, \ \  l = 1, ... , L
\end{equation}

In the above relation, $z_0 \equiv \bs{x}$ and $z_L \equiv \bs{u}$ are the inputs and outputs of the model, with $\sigma^l$s being the nonlinear activation functions and $\circ$ being the compositional construction of the network. Notably, the hyperbolic-tangent is frequently employed as the activation function. 

Consider a quasi-static partial differential operator $\mc{P}$ acting on the solution variable $\bs{u}$, as $\mc{P}\bs{u}(\bs{x}) = f(\bs{x})$. It is noteworthy that $\bs{u}$ is the vector form of the displacement field described as $\bs{u}=(u_r,u_\theta,u_z)$ or $\bs{u}=(u_x,u_y,u_z)$ in cylindrical or Cartesian coordinate systems, respectively. The boundary conditions associated with the solution field can also be expressed versus partial differential operator $\mc{B}$ as $\mc{B}\bs{u}(\partial \bs{x}) = g(\partial\bs{x})$, for $\bs{x}\in\mathbb{R}^d$, where $d$ is the spatial dimension of the problem. Suppose the network parameters are all collected in $\bs{\theta} = \bigcup_{i=0}^L (\mathbf{W}^i, \mathbf{b}^i)$. In the context of PINNs, the network parameters are determined through the minimization of a loss function constructed versus the equations governing the problem on the domain and boundary conditions as 
\begin{equation}
\label{eq:loss}
\mc{L}(\bs{x}; \bs{\theta}) = \sum \lambda_i \mc{L}_i = \lambda_1 \left\| \mc{P} \textbf{u} - f \right\| _{\Omega} + \lambda_2\left\| \mc{B} \textbf{u} - g \right\|_{\partial \Omega} + \dots \,\,\,,
\end{equation} 
with $\mc{L}$ being a loss function, and $\lambda_i$s being a selection of weights associated with each loss term which are determined adoptively throughout the solution process. 

The mean squared error norm is elaborated for the evaluation of the loss function, i.e., $\left\| \circ \right\| = \text{MSE}(\circ)$. In this fashion, the network parameters are determined by means of an optimization problem represented by

\begin{equation}
\label{eq:loss_sum1}
\bs{\theta}^* = \argmin_{\bs{\theta}\in\mathbb{R}^D} \mc{L}(\bs{X}; \bs{\theta})\,\,\,,
\end{equation}
in which $D$ is the total number of trainable parameters, with $\bs{X}\in\mathbb{R}^{n\times d}$ being the set of $n$ collocation points used for the optimization of the loss function. In this study, the construction and training of PINNs is performed by taking advantage of the open-source python API SciANN \citep{haghighat2021sciann}, which is implemented on reputed deep-learning packages TensorFlow \citep{abadi2016tensorflow} and Keras \citep{chollet2015keras}. 

\subsection{Domain-decomposed PINNs for Layered Soils}
\label{S:3-2 (decomposition)}

The primary partial differential equation that governs the deformation of pile-soil system is the equilibrium equation described in section \ref{S:2 (GoverningEq)}. Provided that the material properties of the pile as well as each layer of the surrounding soil could be significantly different, we are dealing with inherent discontinuities within the strain field. Indeed, the variations in material stiffness, that could differ by several orders of magnitude, is responsible for the discontinuity in the derivatives of the displacement field. As such, a domain-decomposed PINN solver is a reasonable choice \cite{kharazmi2021hp, niaki2021physics, jagtap2020conservative, jagtap2021extended}. We, therefore, introduce a series of domain factors to distinguish one material domain from another as

\begin{equation}
\label{eq:domains}
\Bar{\Pi}_\alpha(\bs{x})=\left\{\begin{matrix}1\,\,\,,& \bs{x}\in\Omega_\alpha\\0\,\,\,,&\bs{x}\in\Gamma_\alpha\\-1\,\,\,,&\bs{x}\notin(\Omega_\alpha\cup\Gamma_\alpha)\\\end{matrix}\right.
\end{equation}
in which $\alpha =P,\ S$ for homogeneous domains, and $ \alpha =P, \ S_k\  (k=1,..., N)$ in case of layered formations. In line with the definition of domain factors, it is required to define a distinct neural network for each domain involved. This facilitates the presence of a weak discontinuity in the displacement field across the material interfaces. Hence, the displacement field can be represented over the entire solution domain as

\begin{equation}
\label{eq:discretization}
\begin{gathered}
\begin{aligned}
 u_i  \simeq \mathcal{N}_{u_i}^ P(\bs{x}).\Pi_P(\bs{x})+\sum_{k=1}^{N} \mathcal{N}_{u_i}^ {S_k}(\bs{x}). \Pi_{S_k}(\bs{x})\,\,\,,
\end{aligned}
\end{gathered}
\end{equation}
where $\Pi_\alpha=(\Bar{\Pi}_\alpha+1)/2$. 


As mentioned in section \ref{S:2 (GoverningEq)}, in order to avoid non-physical overlap/separation of the displacement fields associated with different solution domains, displacement constraints need to be imposed along the material interfaces. For the sake of simplicity, it is also stipulated that no-slip may occur along any material interface. As such, the compatibility constraints corresponding to the displacement field across the whole domain can be expressed by

\begin{equation}
\label{eq:contact}
\left\{\begin{array}{l}
                  {N}_{u_i}^\alpha(\bs{x}) - {N}_{u_i}^\beta(\bs{x})=0\,\,\,,
                  \\
                  \\
                 \forall  \bs{x}\in (\Gamma_\alpha\cap\Gamma_\beta)
                \,\,\,\text{, where} \ \alpha \neq \beta
                \end{array}
              \right.
\end{equation}
where $i= (r,\theta,z)$ or $(x,y,z)$ for the Cylindrical and Cartesian coordinates, respectively.


\section{PINNs Solution and Parametric Study}
\label{S:4 (solutions)}

In this section, application of the proposed framework is demonstrated for the solution and parametric study of piles in both homogeneous and layered formations. For this sake, the governing equations presented in section \ref{S:2 (GoverningEq)} are employed for the solution of pile-soil systems under axisymmetric and plane strain conditions. The performance of the framework for parameter identification in layered soils is explored in the final example.


\subsection{Forward Solution of Cylindrical Piles in Homogeneous Soils}

\begin{figure}[!b]
\centering
\begin{subfigure}{0.99\textwidth}
    \centering
    \includegraphics[trim={0 0 0 0},width=0.6\linewidth]{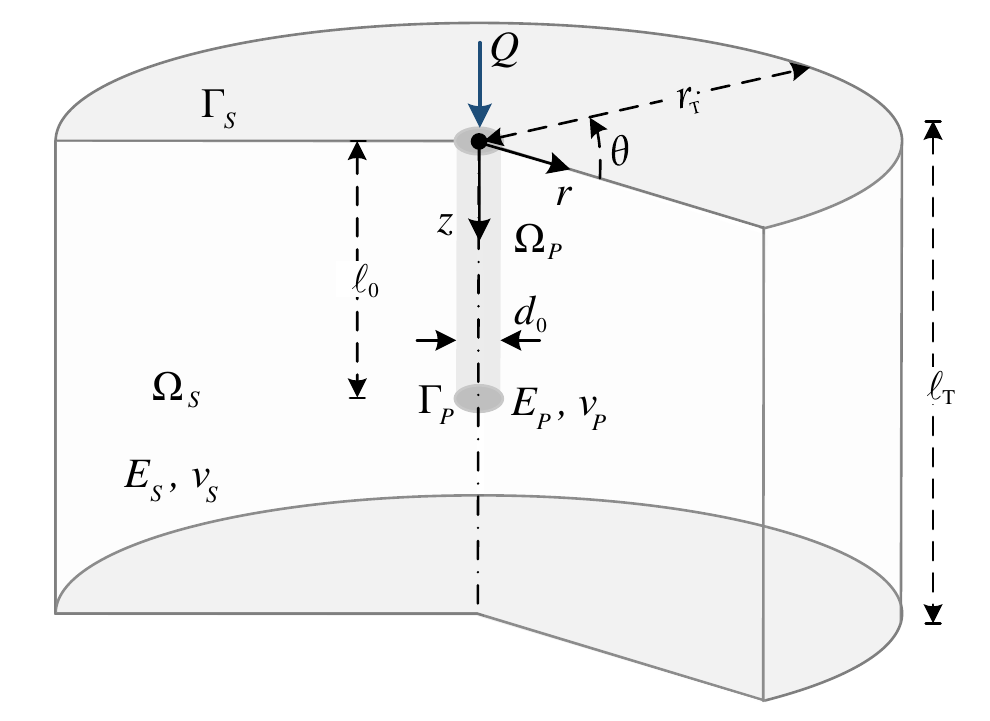}
\end{subfigure}
\caption{The cylindrical pile embedded in homogeneous soil; problem definition and boundary conditions.}
\label{fig:Axis-symmetric pile}
\end{figure}

\begin{figure}[!t]
\centering
\begin{subfigure}{0.48\textwidth}
    \centering
    \includegraphics[trim={0 0 0 0},width=0.99\linewidth]{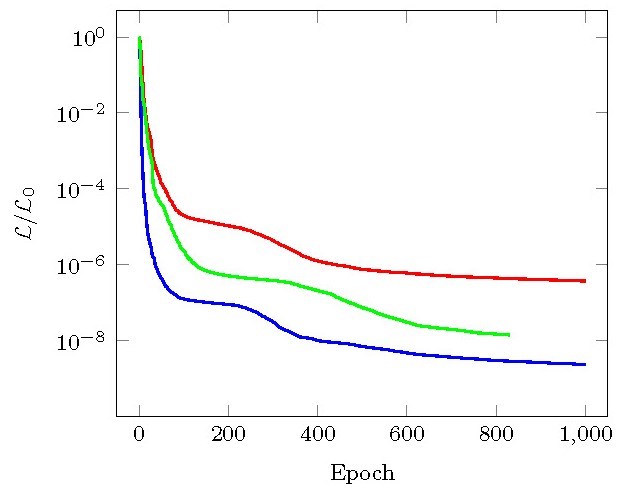}
\end{subfigure}
\begin{subfigure}{0.48\textwidth}
    \centering
    \includegraphics[trim={0 0 0 0},width=0.99\linewidth]{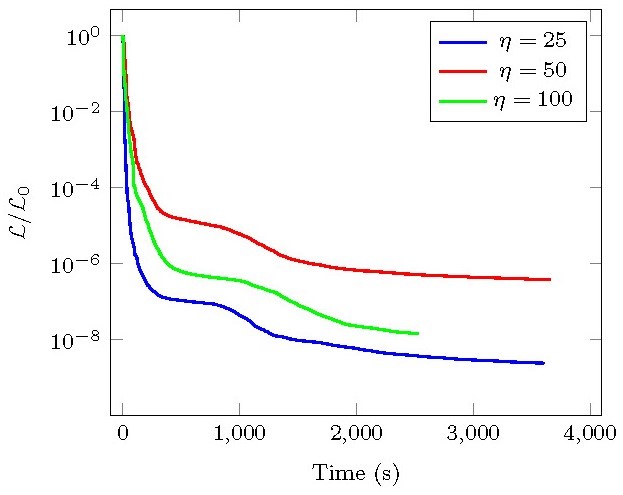}
\end{subfigure}
\caption{The network training history for the cylindrical pile in homogeneous domains.}
\label{fig:axis_pile_loss}
\end{figure}

\begin{figure}[!b]
\centering
\begin{subfigure}{0.32\textwidth}
    \centering
    \includegraphics[trim={0 0 0 0},width=0.99\linewidth]{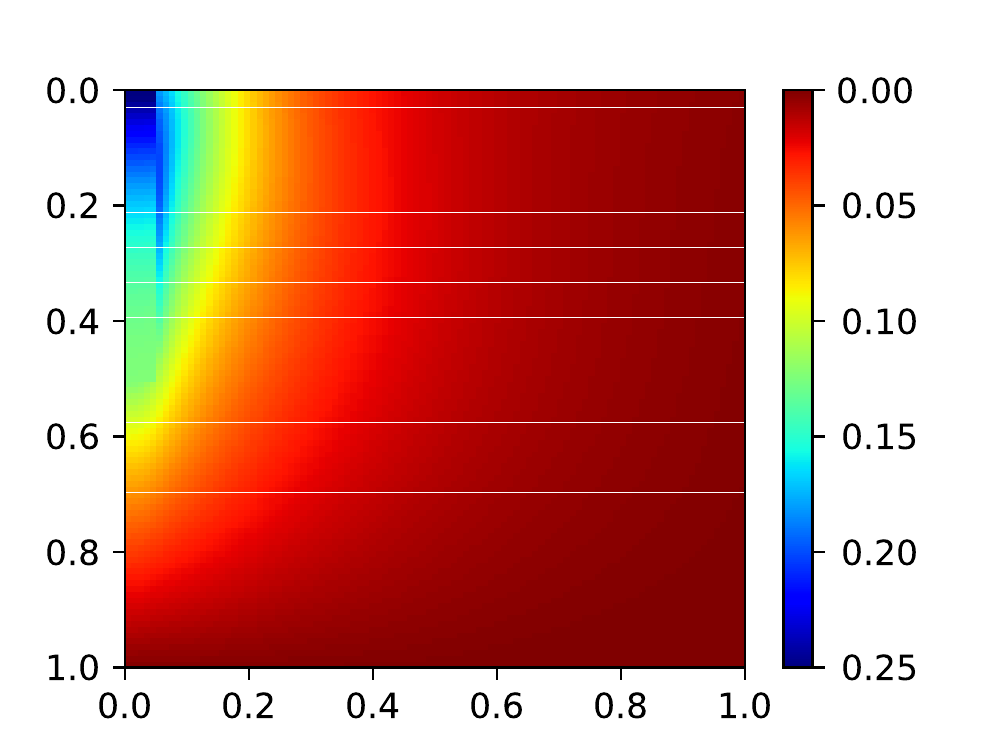}
    \caption{PINN solution for $\eta=25$}
    \label{fig:EX1-a}
\end{subfigure}
\begin{subfigure}{0.32\textwidth}
    \centering
    \includegraphics[trim={0 0 0 0},width=0.99\linewidth]{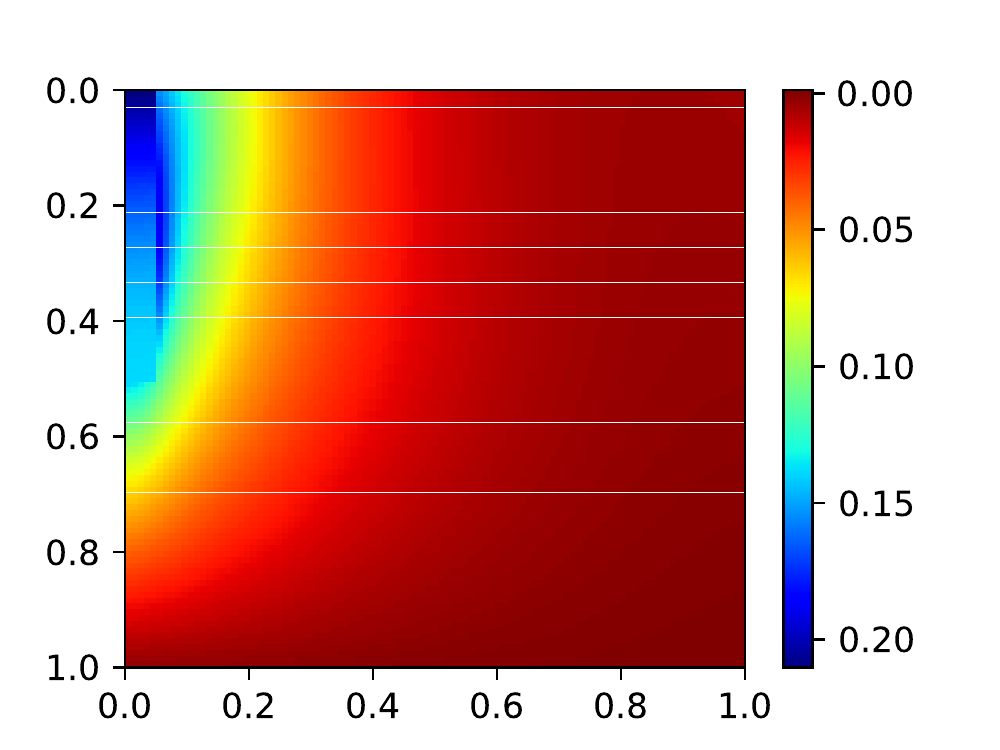}
    \caption{PINN solution for $\eta=50$}
    \label{fig:EX1-b}
\end{subfigure}
\begin{subfigure}{0.32\textwidth}
    \centering
    \includegraphics[trim={0 0 0 0},width=0.99\linewidth]{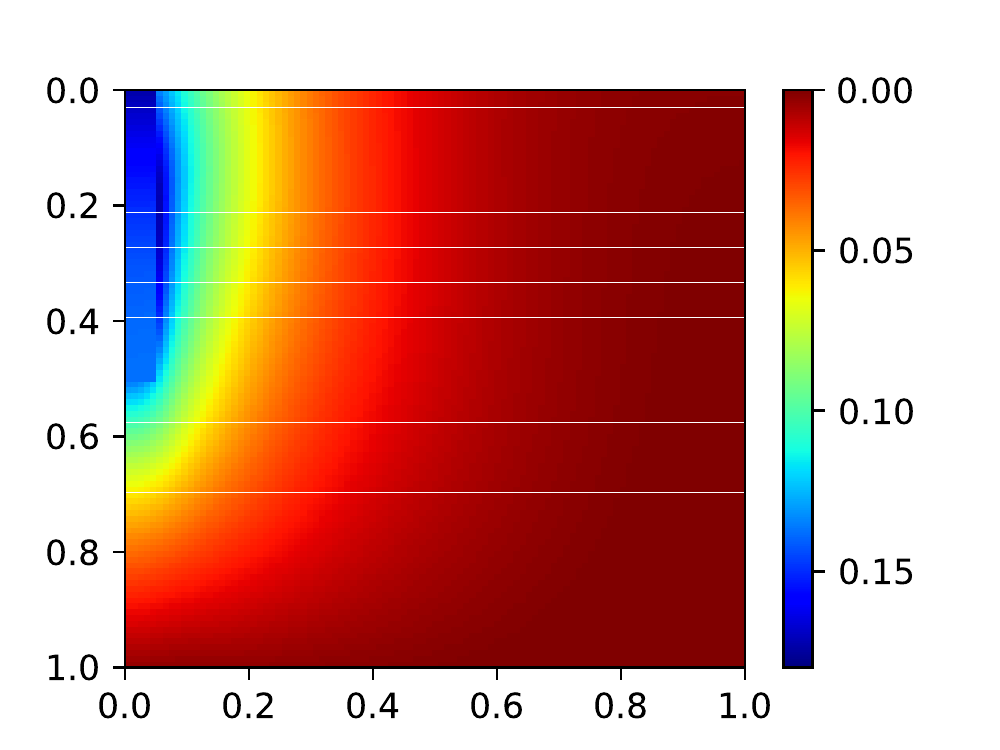}
    \caption{PINN solution for $\eta=100$}
    \label{fig:EX1-c}
\end{subfigure}
\begin{subfigure}{0.32\textwidth}
    \centering
    \includegraphics[trim={0 0 0 0},width=0.8\linewidth]{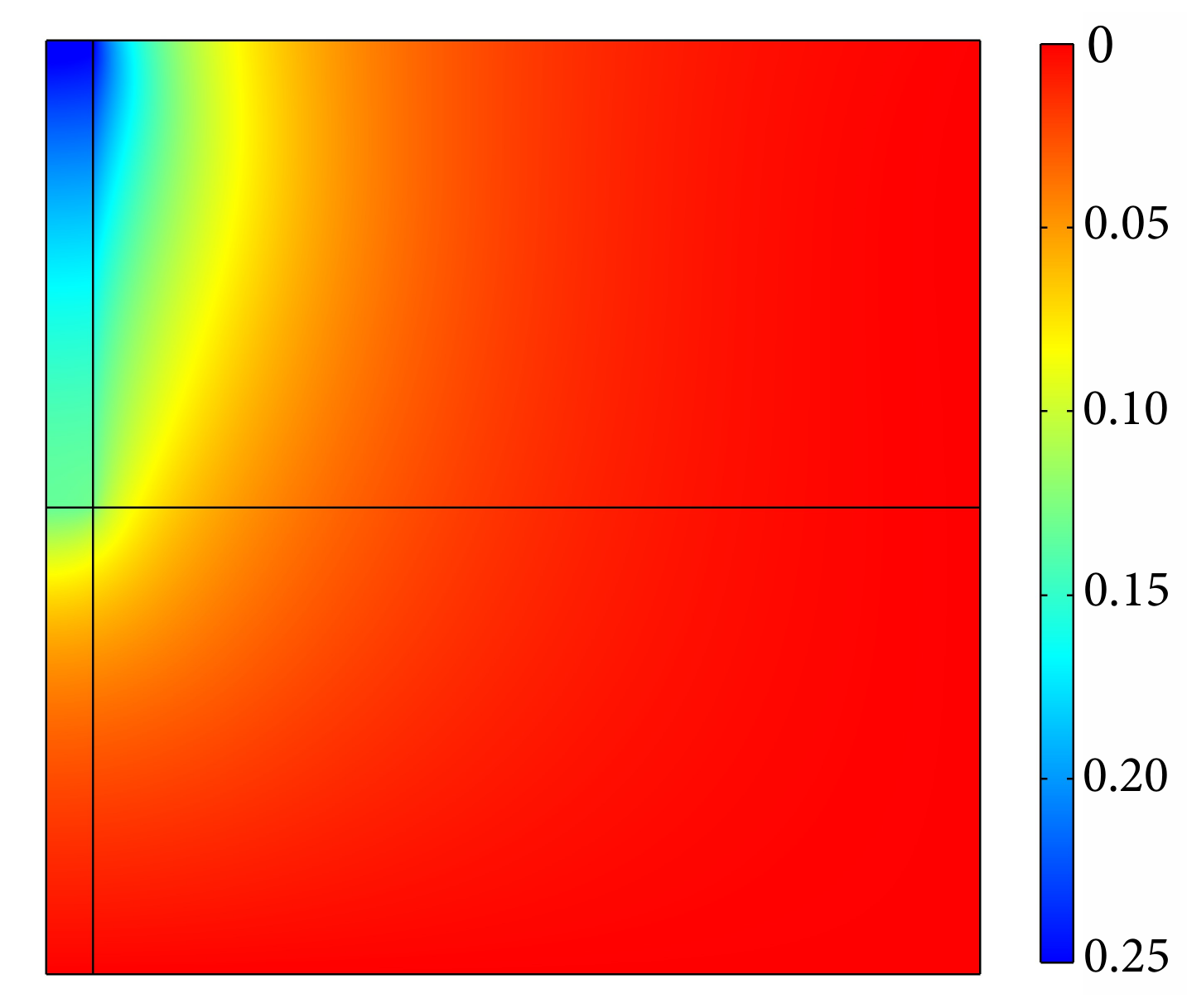}
    \label{fig:EX1-d}
    \caption{Reference solution for $\eta=25$}
\end{subfigure}
\begin{subfigure}{0.32\textwidth}
    \centering
    \includegraphics[trim={0 0 0 0},width=0.8\linewidth]{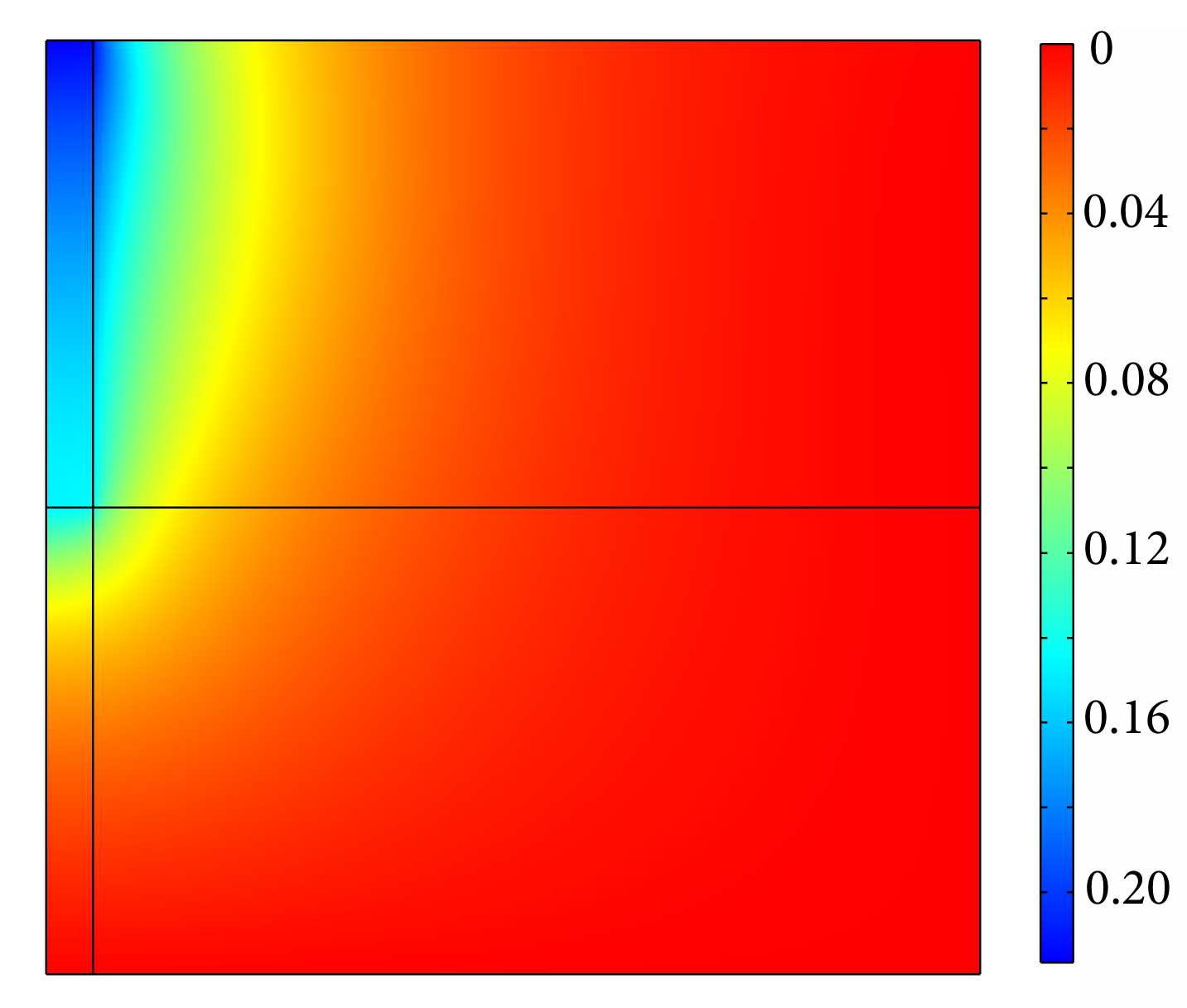}
    \caption{Reference solution for $\eta=50$}
    \label{fig:EX1-e}
\end{subfigure}
\begin{subfigure}{0.32\textwidth}
    \centering
    \includegraphics[trim={0 0 0 0},width=0.8\linewidth]{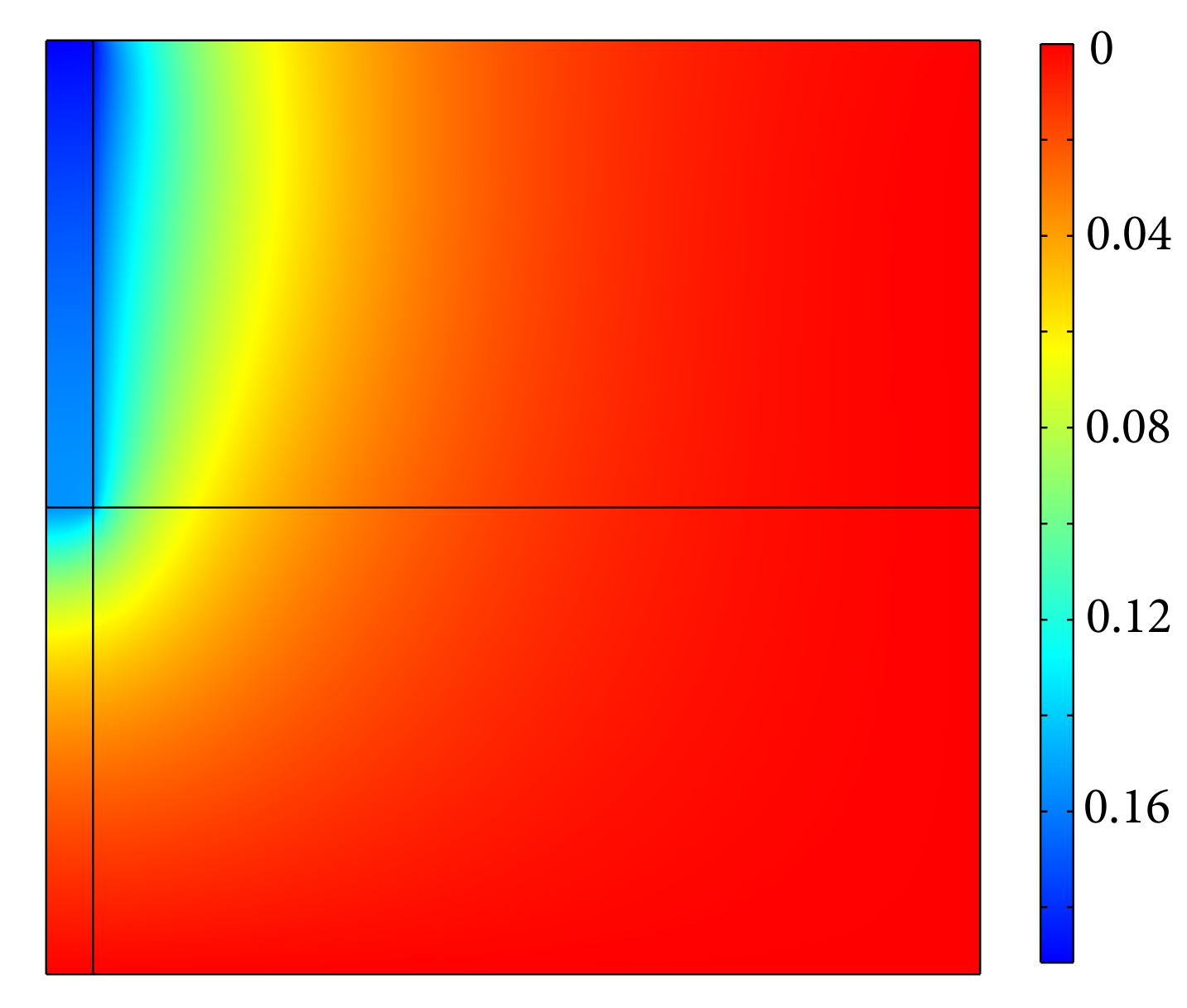}
    \caption{Reference solution for $\eta=100$}
    \label{fig:EX1-f}
\end{subfigure}
\caption{Contours of the normalized vertical displacement $u_z/\ell_T (10^{-2})$ for the cylindrical pile in homogeneous domains.}
\label{fig:axis-pile_contour}
\end{figure}

In this example, the PINNs solution of an axisymmetric cylindrical pile under vertical loading is investigated. The general form of the PDEs governing the deformation of the pile-soil system is given by the Equilibrium Eq.s \eqref{eq:Equilibrium}. Considering the axisymmetricity of the problem, all derivatives with respect to $\theta$ and the terms related to shear stresses exerted in direction $\theta$ are vanished. To avoid the singularity of the governing equations, the equilibrium equations are multiplied by $r$. Thus, the simplified form of the equilibrium equations is expressed as


\begin{equation}
\begin{gathered}
\begin{aligned}
\label{eq:EquilibriumEx1}
  & \frac{\partial }{\partial r}\left( r{{\sigma }^\alpha_{rr}} \right)+r\frac{\partial {{\sigma }^\alpha_{rz}}}{\partial z}-{{{\sigma }^\alpha_{\theta \theta }}} \equiv \mc{P}_{rr}^\alpha \textbf{u}^\alpha = 0\,\,\,, \\ 
  &  \\ 
  & \frac{\partial }{\partial r}\left( r{{\sigma }^\alpha_{rz}} \right)+r\frac{\partial {{\sigma }^\alpha_{zz}}}{\partial z} \equiv \mc{P}_{zz}^\alpha \textbf{u}^\alpha =0\,\,\,. \\ 
\end{aligned}
\end{gathered}
\end{equation}

The strain field is described by Eq. \eqref{eq:StrainDisp}, except that all terms involving $u_\theta$ or derivatives with respect to $\theta$ are vanished. In addition, the singularities due to the presence of the term $1/r$ at the origin (i.e., $r=0$) are relieved through the addition of an small amount $\Bar{\epsilon}$, as $1/r \simeq 1/(r+\Bar{\epsilon})$ (here, $\Bar{\epsilon}=0.001$). Therefore, it follows that

\begin{equation}
\begin{gathered}
\begin{aligned}
\label{eq:StrainDispEx1}
  & {{\varepsilon }_{rr} ^{\alpha}}=\frac{\partial {{u}^\alpha_{r}}}{\partial r}\,\,\,,\,\,\,\,\,\,\,{{\varepsilon }^\alpha_{\theta \theta }}=\frac{{u}^\alpha_{r}}{(r+\Bar{\epsilon})}\,\,\,,\,\,\,\,\,\,\,{{\varepsilon }^\alpha_{zz}}=\frac{\partial {{u}^\alpha_{z}}}{\partial z}\,\,\,, \\ 
 &  \\ 
 & {{\varepsilon }^\alpha_{rz}}=\frac{1}{2}\left( \frac{\partial {{u}^\alpha_{r}}}{\partial z}+\frac{\partial {{u}^\alpha_{z}}}{\partial r} \right)\,\,\,, \\ 
\end{aligned}
\end{gathered}
\end{equation}
where $\alpha=P,\ S_k$. The above equations are accompanied by the constitutive relation given by Eq. \eqref{eq:Constitutive}.

The compatibility constraint required for the interface formed at the intersection of pile and soil is expressed by
\begin{equation}
\label{eq:contact0}
\left\{\begin{array}{ll}
                  u^P_r (\bs{x}) - u^S_r (\bs{x})=0\,\,\,, \\
                  \\
                 u^P_z (\bs{x}) - u^S_z (\bs{x})=0\,\,\,, \\
                  \\
                 \forall  \bs{x}\in (\Gamma_P\cap\Gamma_S)
                \end{array}
              \right.
\end{equation}

As depicted in Fig. \ref{fig:Axis-symmetric pile}, suppose a cylindrical pile with the slenderness ratio of $\ell_0/d_0=5$, which is subject to the vertical loading of $Q=100 \text{ kN}$. The domain consists of a homogeneous soil layer that is extended for the normalized radius of $r_T/d_0=10$ and length of  $\ell_T/\ell_0=2$. The material properties for the soil are assumed as: Young's Modulus of Elasticity, $E_S=100 \text{ MPa}$; Poisson's ratio, $\nu_S=0.25$. The material properties of the pile are: Young's Modulus of Elasticity, $E_P=2.5, 5, 10 \text{ GPa}$; Poisson's ratio, $\nu_P=0.25$. In this fashion, the problem is studied for the stiffness ratios $\eta=E_P/E_S=25,\ 50\, 100$.

In order to perform the PINNs solution, as explained in section \ref{S:3 (PINN)}, we need to use multiple neural networks proportional to the number of materials existing throughout the entire domain. As such, two distinct neural networks are introduced as per component of the displacement field as
\begin{equation}\label{eq:NN5}
\begin{split}
u_r^P &\simeq \mathcal{N}_{u_r}^P(r,z)\,\,\,, \ \ \ u_z^P \simeq \mathcal{N}_{u_z}^P(r,z)\,\,\,, \\
u_r^S &\simeq \mathcal{N}_{u_r}^S(r,z)\,\,\,, \ \ \  u_z^S \simeq \mathcal{N}_{u_z}^S(r,z)\,\,\,. \\
\end{split}
\end{equation}

4 hidden layers with 20 neurons in each layer are considered in all the cases. Hyperbolic-tangent is also used as the activation function. The physics-informed loss terms of the total cost function are expressed as

\begin{equation}
\begin{split}
\label{eq:loss_sum2}
&\mc{L}_T = \mathcal{L}_\Omega + \mathcal{L}_{\Gamma_\text{B.C.}} +\mathcal{L}_{\Gamma_\text{Cont}}\,\,\,,
\\
\\
&\mathcal{L}_\Omega = \lambda_1 \left\|\mc{P}_{rr}^P \textbf{u}^P\right\|_{\text{on }\Omega_P}+ \lambda_2 \left\|\mc{P}_{rr}^S\textbf{u}^S\right\|_{\text{on }\Omega_S}
\\
&\ \ \ \ +  \lambda_3 \left\|\mc{P}_{zz}^P \textbf{u}^P\right\|_{\text{on }\Omega_P}+\lambda_4 \left\|\mc{P}_{zz}^S \textbf{u}^S\right\|_{\text{on }\Omega_S}\,\,\,,
\\
\\
&\mathcal{L}_{\Gamma_\text{B.C.}} =\lambda_5 \left\| \mc{B}_{rr}^P \textbf{u}^P - g_{rr}^P \right\|_{\text{on } \Gamma_P \setminus \Gamma_S} + \lambda_6 \left\| \mc{B}_{rr}^S \textbf{u}^S - g_{rr}^S \right\|_{\text{on } \Gamma_S \setminus \Gamma_P}
\\
&\ \ \ \ \ \ \ \ +\lambda_7 \left\| \mc{B}_{zz}^P \textbf{u}^P - g_{zz}^P \right\|_{\text{on } \Gamma_P \setminus \Gamma_S}+\lambda_8 \left\| \mc{B}_{zz}^S \textbf{u}^S - g_{zz}^S \right\|_{\text{on } \Gamma_S \setminus \Gamma_P}\,\,\,,
\\
\\
&\mathcal{L}_{\Gamma_\text{Cont}} =\lambda_9 \left\| \textbf{u}^P - \textbf{u}^S \right\|_{\text{on } _{\Gamma_P \cap \Gamma_S}}+\lambda_{10} \left\| \textbf{t}^P - \textbf{t}^S \right\|_{\text{on } _{\Gamma_P \cap \Gamma_S}}\,\,\,,\\
\end{split}
\end{equation}
where $\mc{P}^\alpha$, $\mc{B}^\alpha$ and $g^\alpha$ are differential operators associated with the equilibrium equations, boundary conditions, and preassigned boundary values of the problem, respectively, in the directions $r$ and $z$.  

The training is performed by using 3000 sampling points over each of domains $\Omega_P$ and $\Omega_S$ (i.e., 6000 in total), whereas at least $50\%$ of the sampling points are clustered over the boundaries $\Gamma_P$ and $\Gamma_S$. The use of non-uniform sampling grid ensures that the complex boundary conditions introduced along the pile-soil interface as well as the external boundaries are properly satisfied. Furthermore, Neural Tangent Kernel (NTK) adaptive weighting is employed to determine $\lambda$s, which guarantees all loss terms are calibrated proportionally throughout the training process \cite{wang2022and}. Here, the Adam optimization scheme is adopted for the training with the learning rate of $0.003$.

In Fig. \ref{fig:axis_pile_loss}, the evolution of normalized loss versus epochs and time is illustrated. Evidently, in all cases, the designated architecture converges rapidly to the relative error of $10^{-6}$ within less than $1000$ epochs. Meanwhile, the training process is accomplished relatively fast (total duration $\approx 3000 \text{ s}$). Contours of the normalized vertical displacement field $u_z/\ell_T$ is shown for all the cases in Fig. \ref{fig:axis-pile_contour}. Furthermore,  a reference solution obtained using the FEM software COMSOL Multiphysics \cite{multiphysics1998introduction} is presented for the sake of comparison. An excellent agreement is observed between the PINNs solution and the results of the FEM analysis. This demonstrates the validity of the proposed framework for the analysis of pile-soil systems under the axisymmetric condition.

\subsection{Forward Solution of Piles in Homogeneous Soils Under Plane Strain Condition}

\begin{figure}[!t]
\centering
\begin{subfigure}{0.99\textwidth}
    \centering
    \includegraphics[trim={0 0 0 0},width=0.6\linewidth]{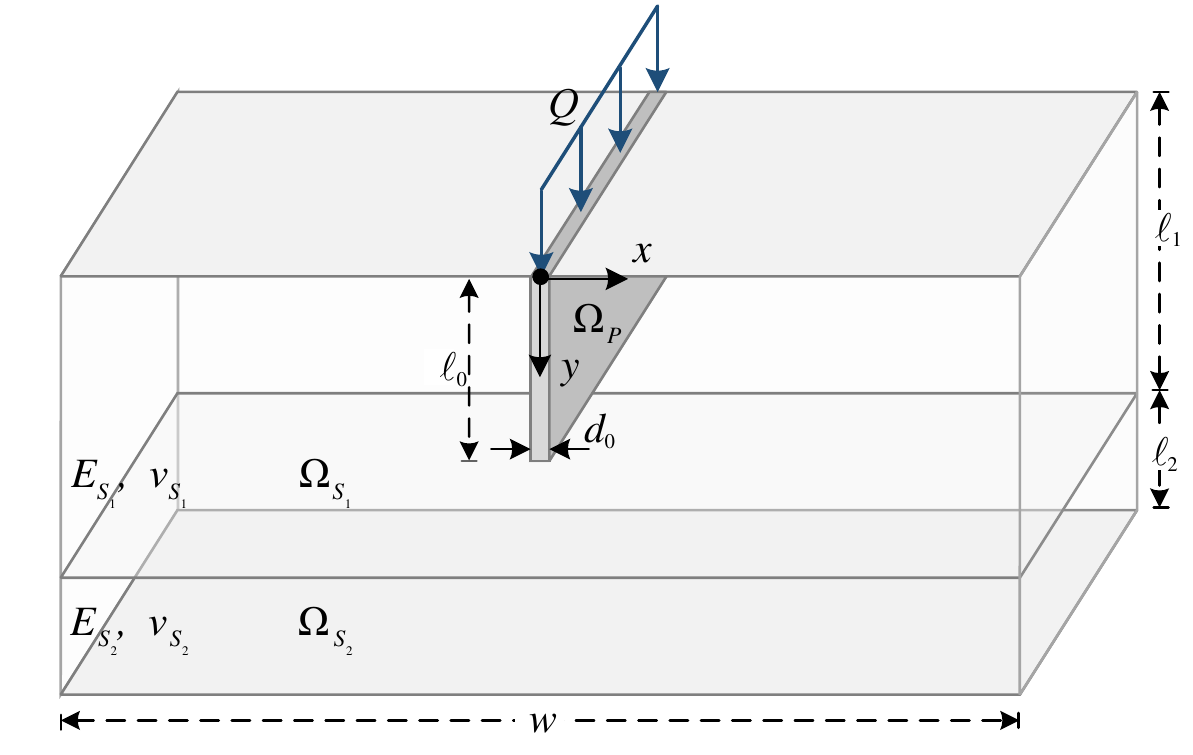}
\end{subfigure}
\caption{The sheet-pile wall in layered formation; problem definition and boundary conditions.}
\label{fig:Cartesian pile}
\end{figure}

This example investigates the PINN solution of a sheet-pile wall subject to vertical loading in homogeneous soils. The response of the wall is governed by the equilibrium Eq. \eqref{GovEqCar} in Cartesian coordinate systems. However, as sheet-piles are relatively long structural members in nature, simplified plane strain description of the equilibrium equation is typically applied for the analysis of their response in soil medium. In this respect, all derivatives with respect to $y$ are vanished. In the absence of body forces, the equilibrium equation under the plane-strain condition can be described as
 

\begin{equation}
\begin{gathered}
\begin{aligned}
\label{GovEqCarEx2}
\begin{aligned}
& \frac{\partial \sigma^\alpha _{xx}}{\partial x}+\frac{\partial\sigma^\alpha _{zx}}{\partial z}\equiv \mc{P}_{xx}^\alpha \textbf{u}^\alpha = 0\,\,\,, \\ \\
& \frac{\partial \sigma^\alpha _{xz}}{\partial x}+\frac{\partial\sigma^\alpha _{zz}}{\partial z}\equiv \mc{P}_{zz}^\alpha \textbf{u}^\alpha = 0\,\,\,.
\end{aligned}
\end{aligned}
\end{gathered}
\end{equation}

In the plane-strain regime, the strain terms manifest in Eq. \eqref{GovEqCar} are further simplified as

\begin{equation}
\begin{gathered}
\label{GovEqCarExample2}
\begin{aligned}
\varepsilon^\alpha _{xx}=\frac{\partial u_{x}^\alpha}{\partial x}\,\,\,, \ \  \varepsilon^\alpha _{zz}=\frac{\partial u_{z}^\alpha}{\partial z}\,\,\,, \ \ \varepsilon^\alpha_{xz}=\varepsilon^\alpha _{zx}=\frac{1}{2} \left( \frac{\partial u_{x}^\alpha}{\partial z}+\frac{\partial u_{z}^\alpha}{\partial x} \right)\,\,\,,
\end{aligned}
\end{gathered}
\end{equation}
where $\alpha=P, \ S_k$. Here, the compatibility constraint for the material interface of pile-soil system is expressed as
\begin{equation}
\label{eq:contact2}
\left\{\begin{array}{ll}
                  u^P_x (\bs{x}) - u^S_x (\bs{x})=0\,\,\,, \\
                  \\
                 u^P_z (\bs{x}) - u^S_z (\bs{x})=0\,\,\,, \\
                  \\
                 \forall  \bs{x}\in (\Gamma_P\cap\Gamma_S)
                \end{array}
              \right.
\end{equation}

Consider a sheet-pile wall with the slenderness ratio of $\ell_0/d_0=5$ subject to a vertical line load of $Q=10,000 \text{ kN/m}$, as shown in Fig. \ref{fig:Cartesian pile}. In this example, it is assumed that the surrounding soil is homogeneous with the depth ratio of $\ell_1/\ell_0=2$. The stiffness ratios $\eta=E_P/E_S=10,\ 25,\ 50$ with $E_P=5 \text{ GPa}$ and Poisson's ratio $\nu_S=\nu_P=0.25$ are considered in this problem. The PINNs solution of this problem is performed by using the below set of neural networks


\begin{equation}\label{eq:NN6}
\begin{split}
u_x^P &\simeq \mathcal{N}_{u_x}^P(x,z)\,\,\,, \ \ \ u_z^P \simeq \mathcal{N}_{u_z}^P(x,z)\,\,\,, \\
u_x^S &\simeq \mathcal{N}_{u_x}^S(x,z)\,\,\,, \ \ \ u_z^S \simeq \mathcal{N}_{u_z}^S(x,z)\,\,\,. \\
\end{split}
\end{equation}
The architecture of all neural networks consists of 4 hidden layers with 20 neurons each, where hyperbolic-tangent is used as the activation function. The loss terms of the total cost function in here are defined identically to the previous example (see Eq. \eqref{eq:loss_sum2}), except that the indices $r$ are now replaced by $x$. The training is performed by means of 6000 sampling points, with 3000 points assumed for each of $\Omega_P$ and $\Omega_S$ domains. NTK adaptive weighting is applied for the training in conjunction with a learning rate of $0.003$.

\begin{figure}[!t]
\centering
\begin{subfigure}{0.48\textwidth}
    \centering
    \includegraphics[trim={0 0 0 0},width=0.99\linewidth]{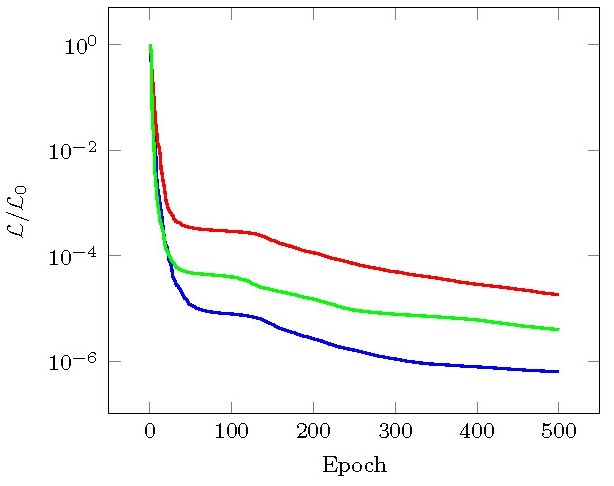}
\end{subfigure}
\begin{subfigure}{0.48\textwidth}
    \centering
    \includegraphics[trim={0 0 0 0},width=0.99\linewidth]{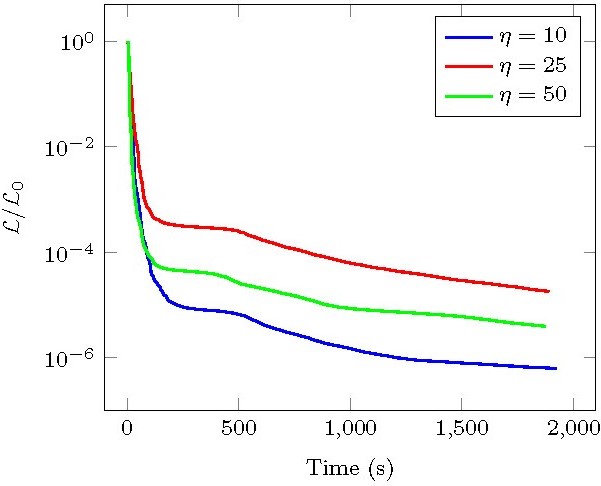}
\end{subfigure}
\caption{The network training history for the sheet-pile wall in homogeneous soils.}
\label{fig:Cartesian_loss}
\end{figure}

\begin{figure}[!b]
\centering
\begin{subfigure}{0.32\textwidth}
    \centering
    \includegraphics[trim={0 0 0 0},width=0.97\linewidth]{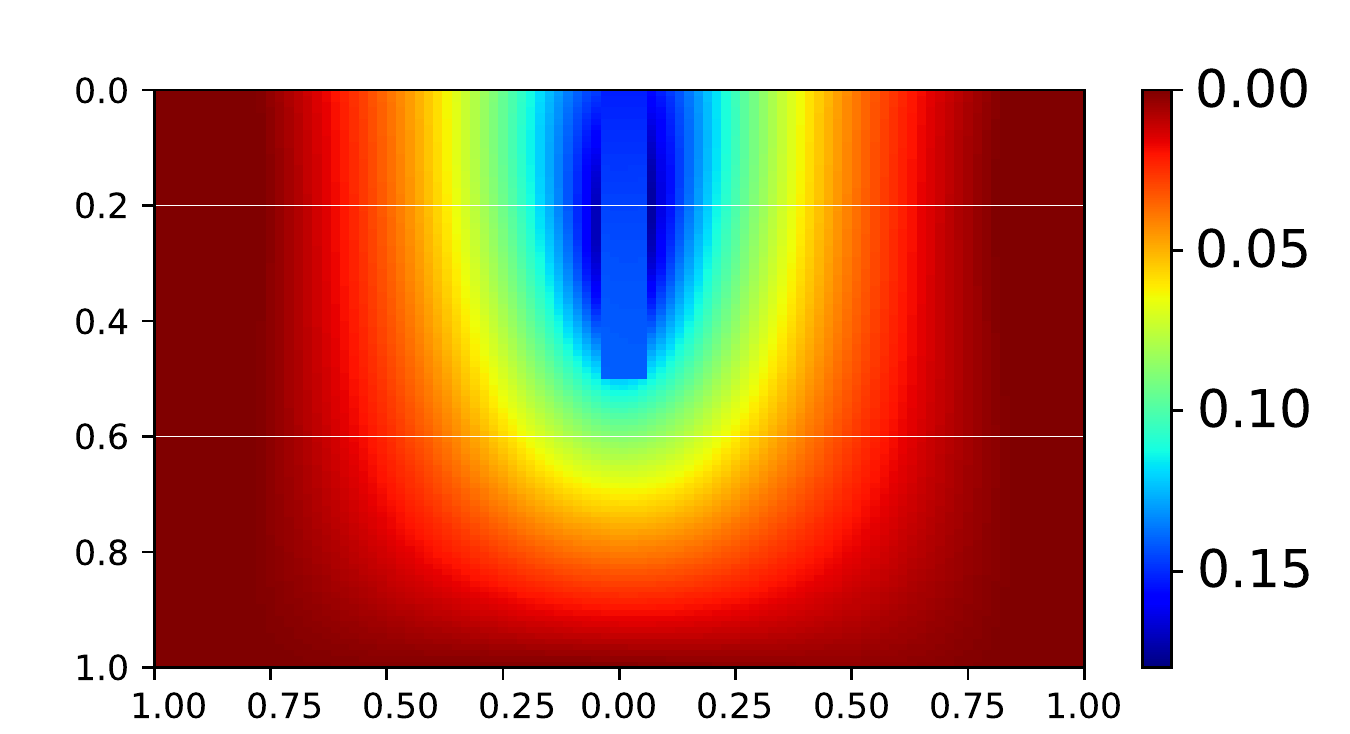}
    \caption{PINN solution for $\eta=10$}
    \label{fig:EX2-a}
\end{subfigure}
\begin{subfigure}{0.32\textwidth}
    \centering
    \includegraphics[trim={0 0 0 0},width=0.97\linewidth]{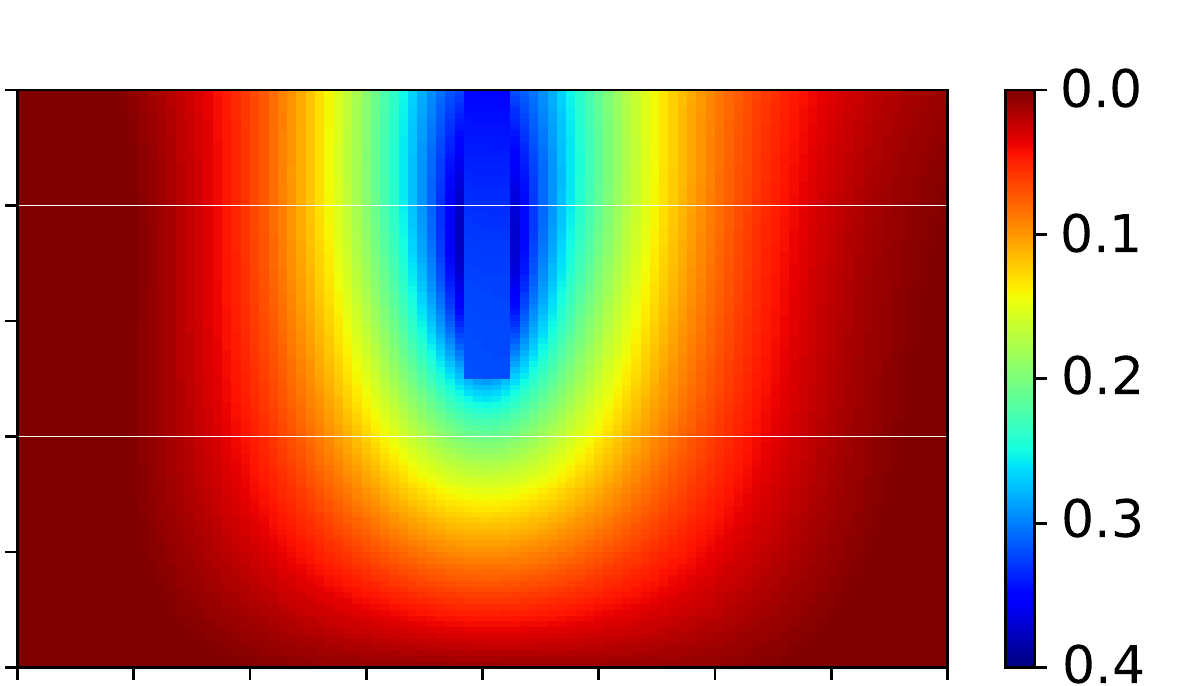}
    \caption{PINN solution for $\eta=25$}
    \label{fig:EX2-b}
\end{subfigure}
\begin{subfigure}{0.32\textwidth}
    \centering
    \includegraphics[trim={0 0 0 0},width=0.97\linewidth]{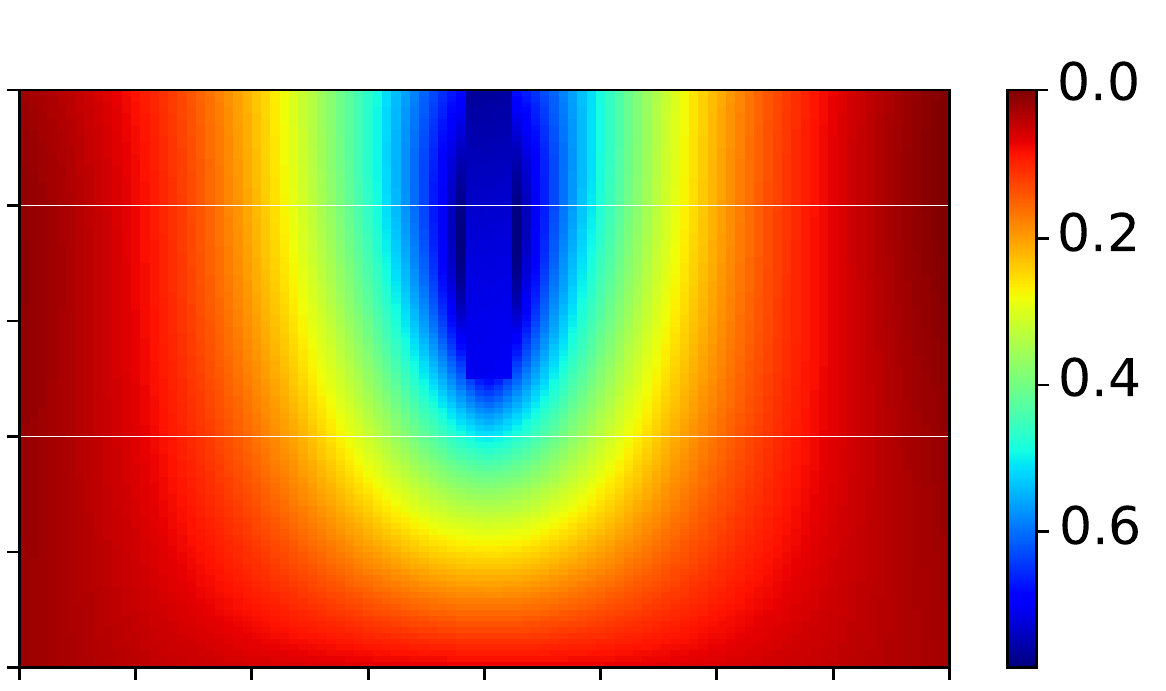}
    \caption{PINN solution for $\eta=50$}
    \label{fig:EX2-c}
\end{subfigure}
\begin{subfigure}{0.32\textwidth}
    \centering
    \includegraphics[trim={0 0 0 0},width=0.93\linewidth]{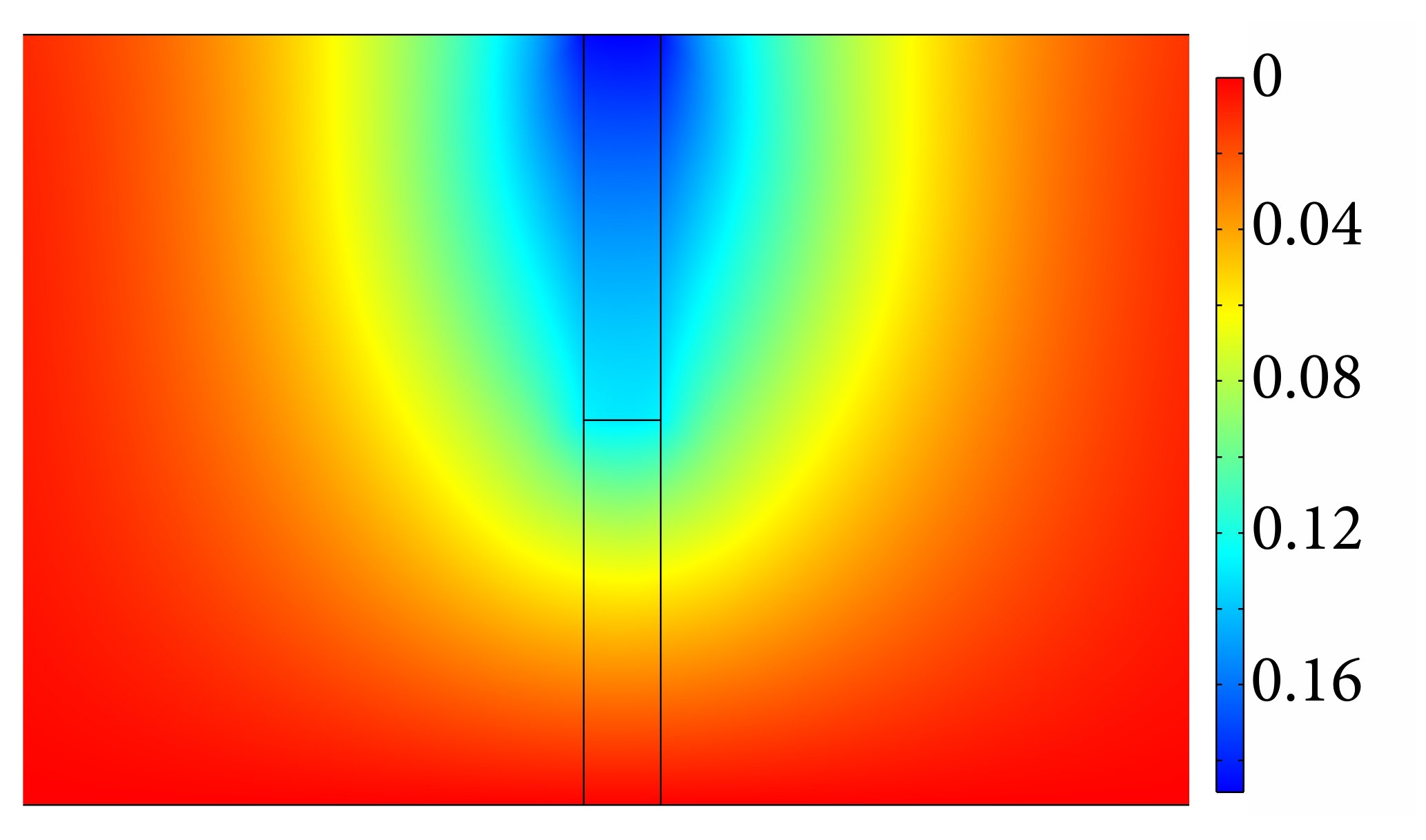}
    \label{fig:EX2-d}
    \caption{Reference solution for $\eta=10$}
\end{subfigure}
\begin{subfigure}{0.32\textwidth}
    \centering
    \includegraphics[trim={0 0 0 0},width=0.93\linewidth]{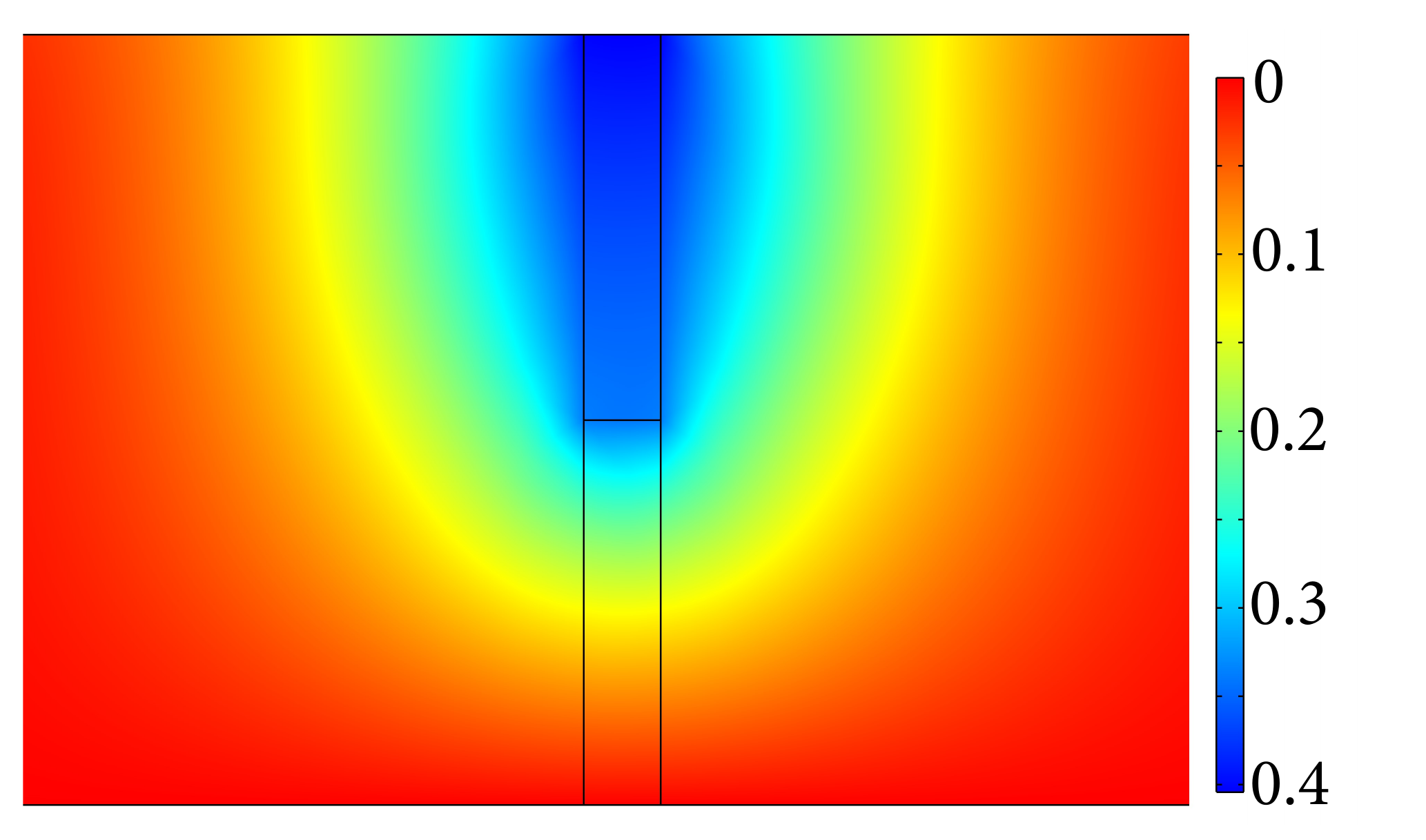}
    \caption{Reference solution for $\eta=25$}
    \label{fig:EX2-e}
\end{subfigure}
\begin{subfigure}{0.32\textwidth}
    \centering
    \includegraphics[trim={-0 0 0 0},width=0.93\linewidth]{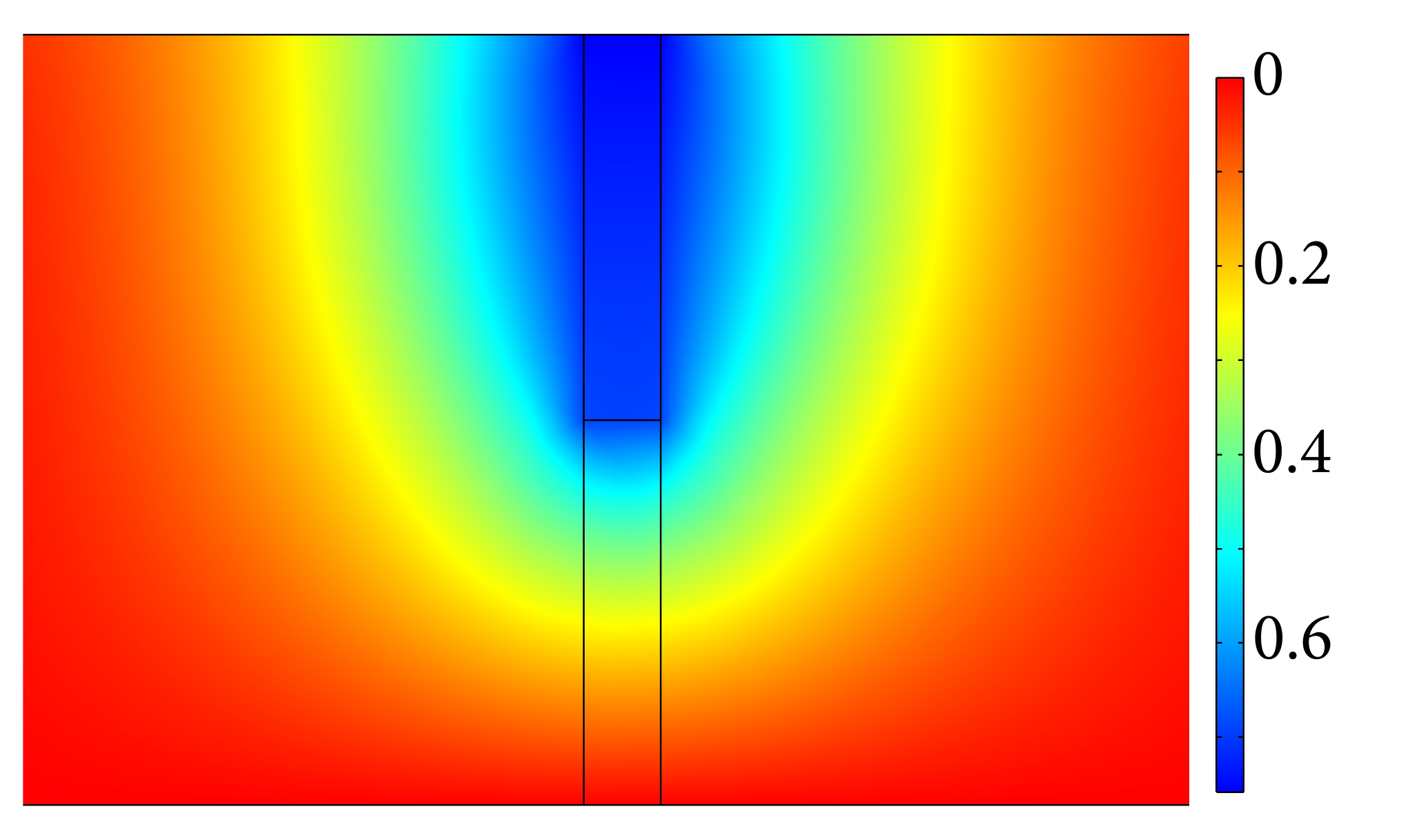}
    \caption{Reference solution for $\eta=50$}
    \label{fig:EX2-f}
\end{subfigure}
\caption{Contours of the normalized vertical displacement $u_z/\ell_1(10^{-2})$ for the sheet-pile wall in homogeneous domains (forward solution).}
\label{fig:ForwardCar}
\end{figure}

The training history of the normalized loss versus epochs and time is reported in Fig. \ref{fig:Cartesian_loss}. As can be seen, the loss function has immediately reached below the relative error norm of $10^{-5}$ within $500$ epochs. The improved performance in terms of convergence rate in comparison to the previous example is attributed to the increased simplicity of the governing equations in the Cartesian system of coordinates. Finally, in Fig. \ref{fig:ForwardCar}, contours of the normalized vertical displacement field $u_{z}/\ell_1$ is presented for all the cases considered based on the PINNs solution and a reference FEM using COMSOL Muiltiphysics. The excellent agreement between the PINNs results and the reference solution indicates the robustness of the extended framework in the study of pile-soil systems under the plane strain condition.

\subsection{Identification of Material Parameters by Inverse Analysis}

\begin{figure}[!t]
\centering
\begin{subfigure}{0.48\textwidth}
    \centering
    \includegraphics[trim={0 0 0 0},width=0.99\linewidth]{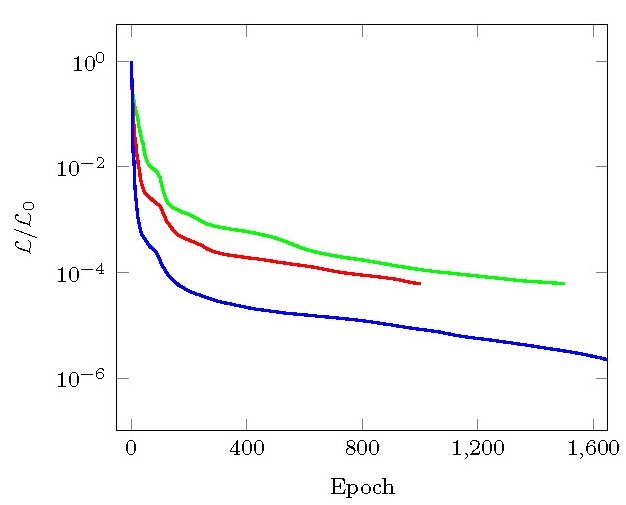}
\end{subfigure}
\begin{subfigure}{0.48\textwidth}
    \centering
    \includegraphics[trim={0 0 0 0},width=0.99\linewidth]{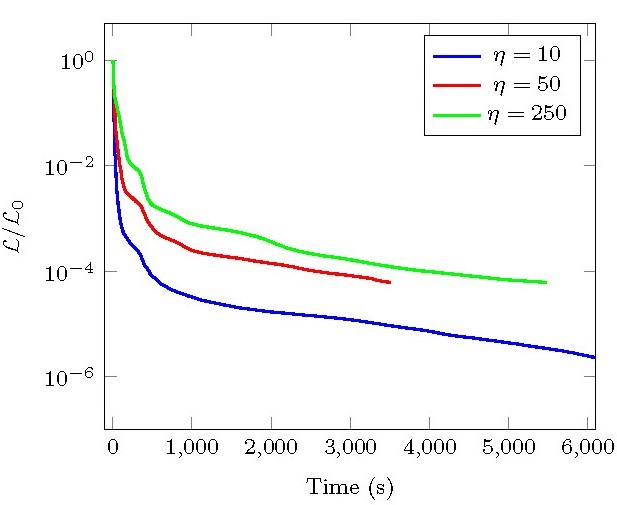}
\end{subfigure}
\caption{The network training history for the inverse analysis of soil-pile interaction in homogeneous domains.}
\label{fig:loss_inversion1}
\end{figure}

\begin{figure}[!b]
\centering
\begin{subfigure}{0.32\textwidth}
    \centering
    \includegraphics[trim={0 0 0 0},width=0.90\linewidth]{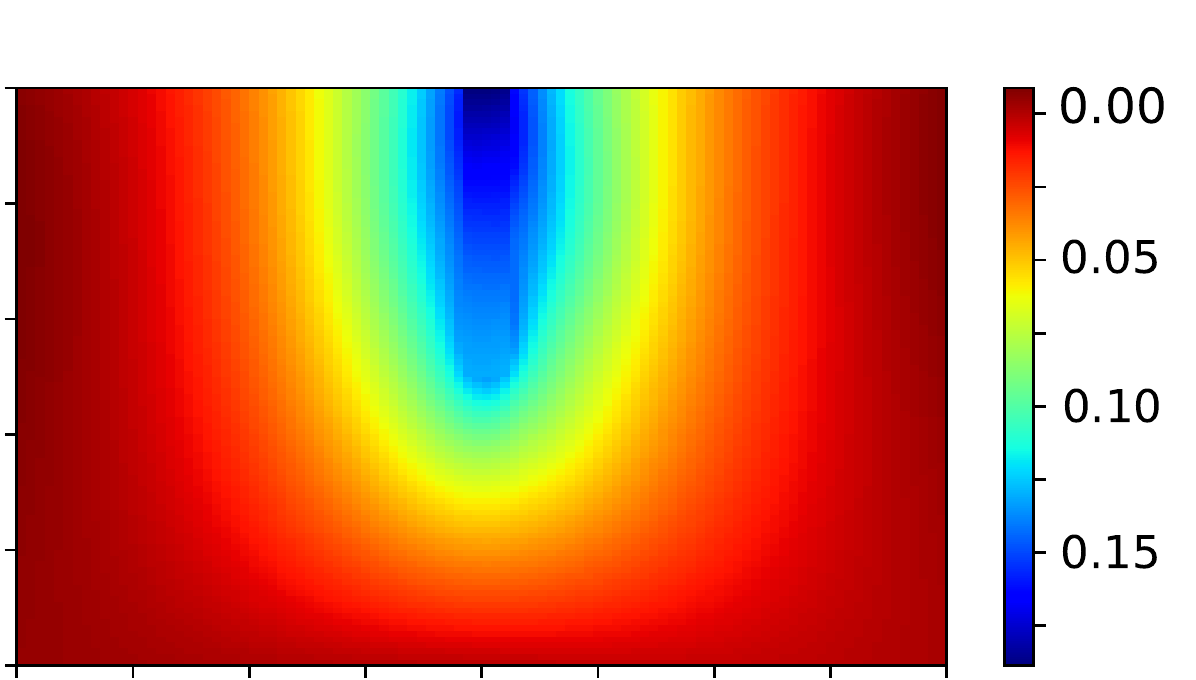}
    \caption{PINN solution for $\eta=10$}
    \label{fig:EX3-a}
\end{subfigure}
\begin{subfigure}{0.32\textwidth}
    \centering
    \includegraphics[trim={0 0 0 0},width=0.87\linewidth]{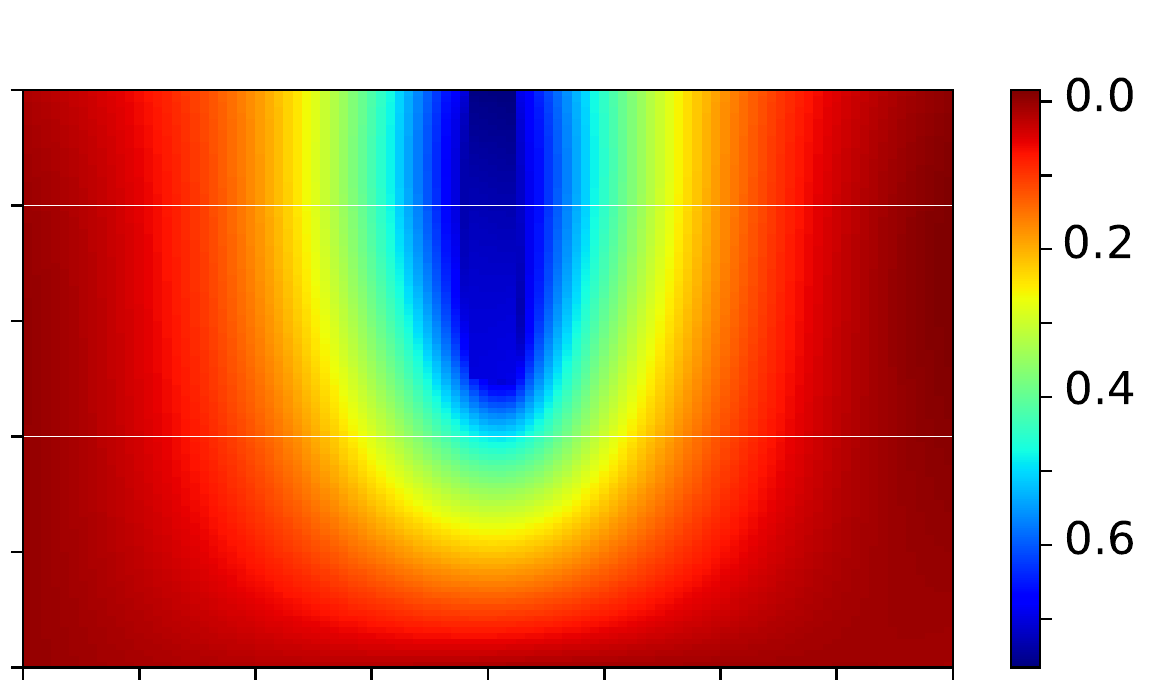}
    \caption{PINN solution for $\eta=50$}
    \label{fig:EX3-b}
\end{subfigure}
\begin{subfigure}{0.3\textwidth}
    \centering
    \includegraphics[trim={0 0 0 0},width=0.90\linewidth]{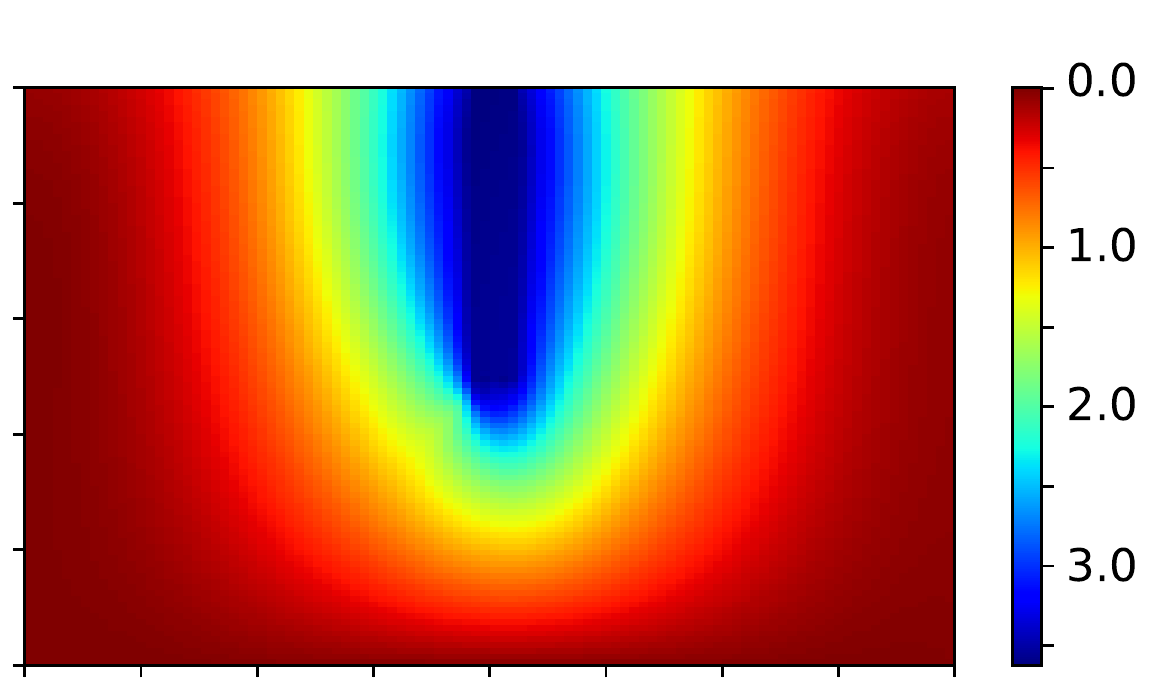}
    \caption{PINN solution for $\eta=250$}
    \label{fig:EX3-c}
\end{subfigure}
\begin{subfigure}{0.32\textwidth}
    \centering
    \includegraphics[trim={-0 0 0 0},width=0.93\linewidth]{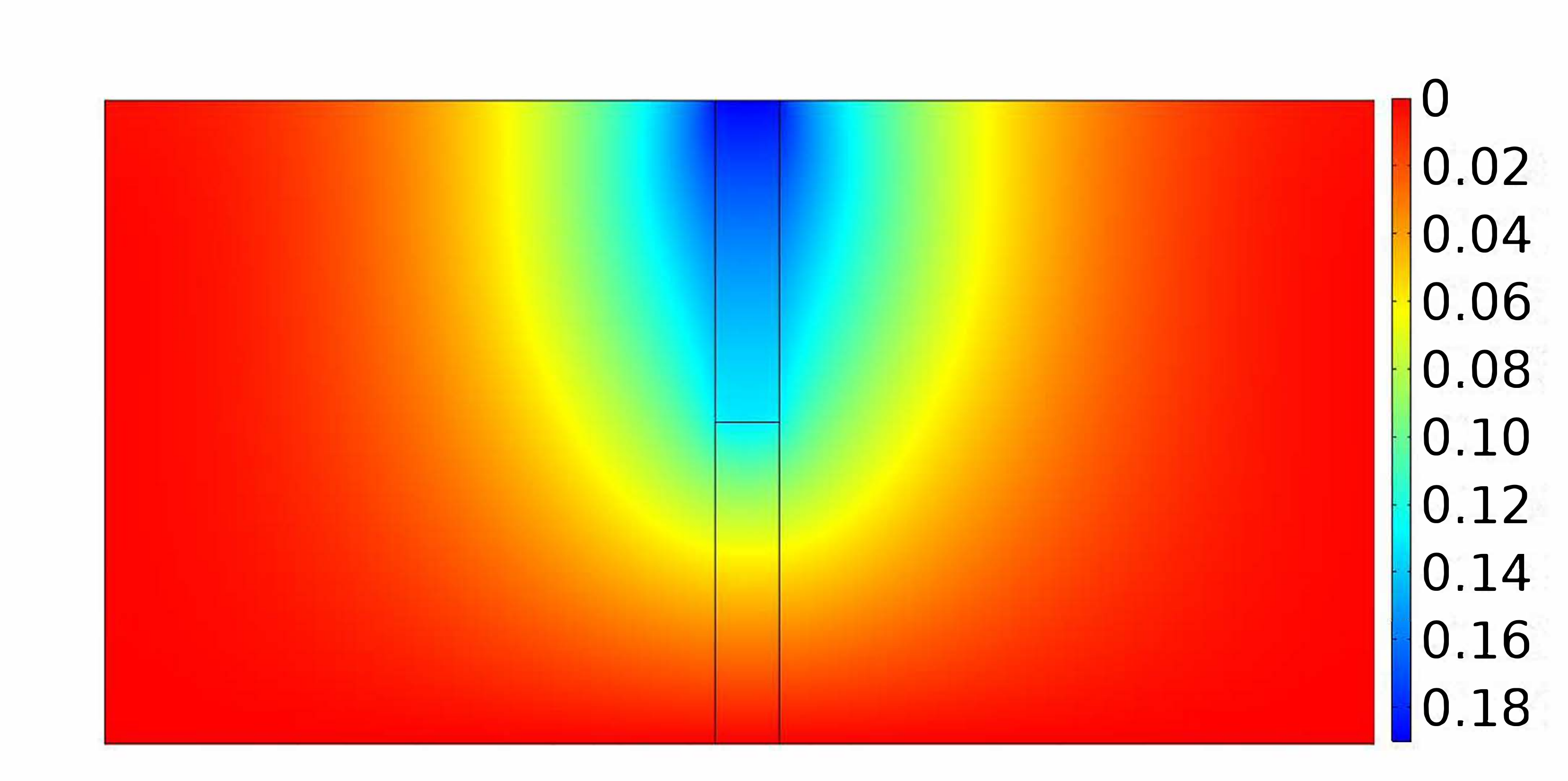}
    \caption{Reference solution for $\eta=10$}
    \label{fig:EX3-d}
\end{subfigure}
\begin{subfigure}{0.32\textwidth}
    \centering
    \includegraphics[trim={0 0 0 0},width=0.93\linewidth]{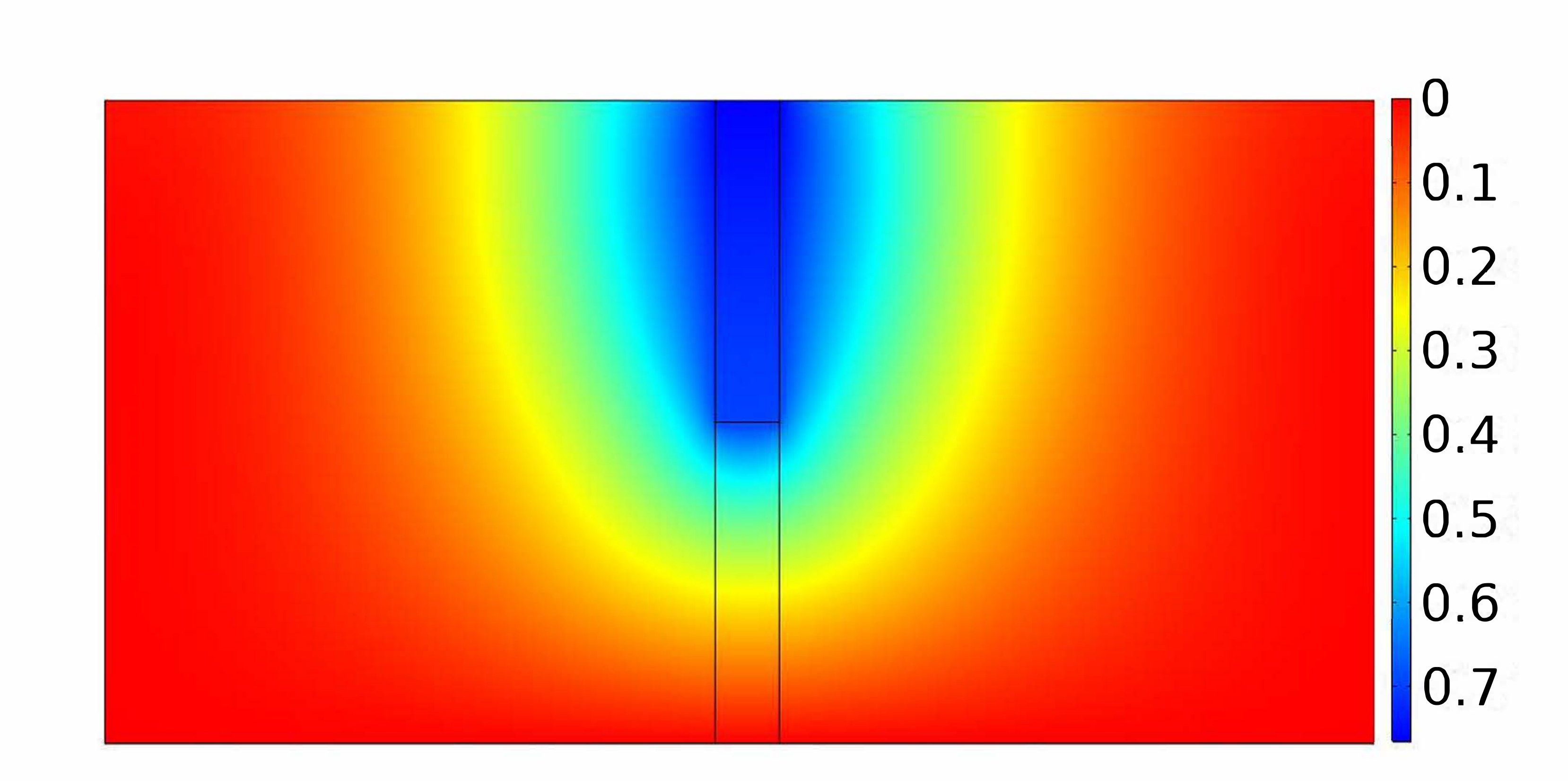}
    \caption{Reference solution for $\eta=50$}
    \label{fig:EX3-e}
\end{subfigure}
\begin{subfigure}{0.32\textwidth}
    \centering
    \includegraphics[trim={0 0 0 0},width=0.93\linewidth]{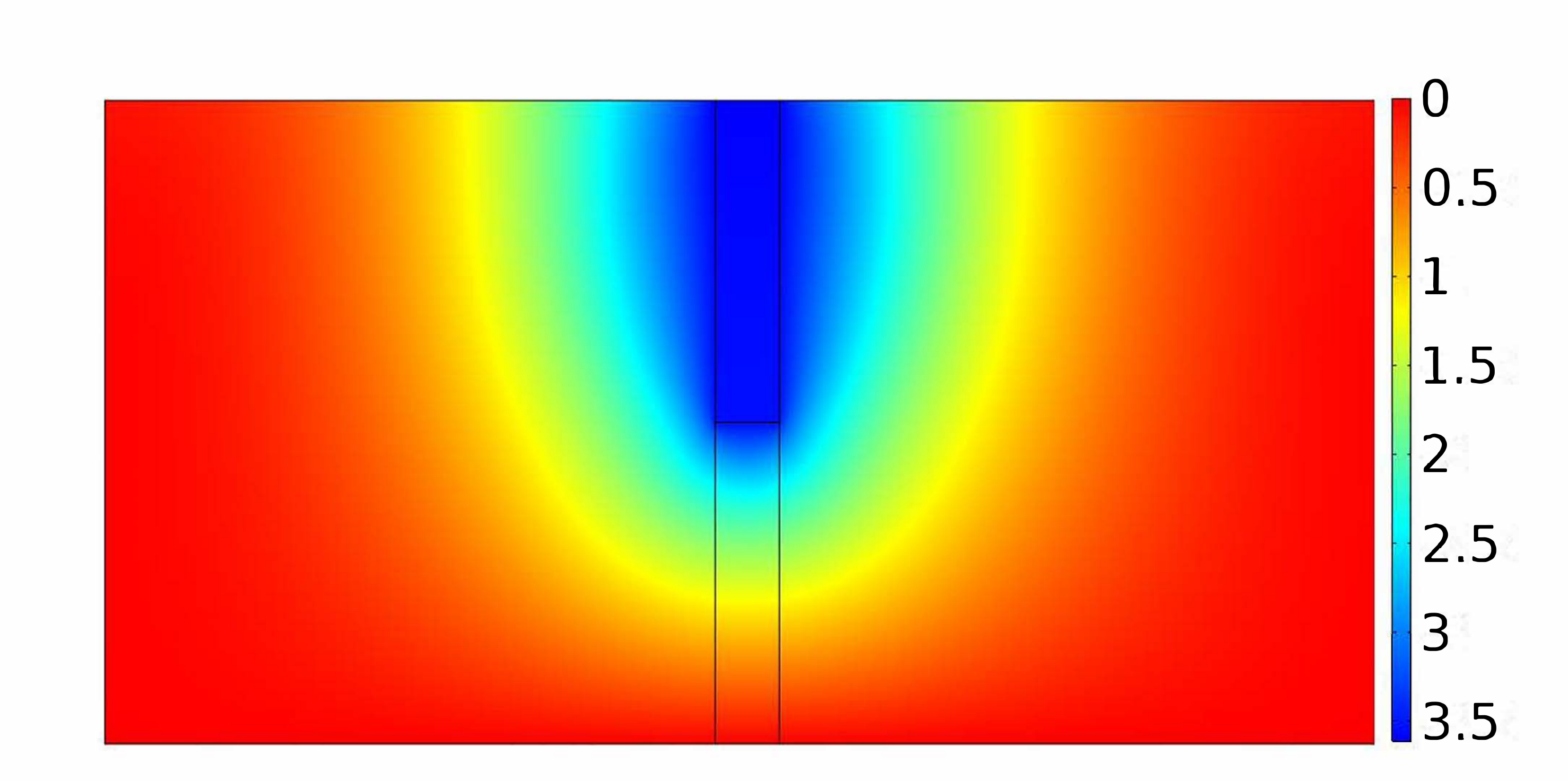}
    \label{fig:EX3-f}
    \caption{Reference solution for $\eta=250$}
\end{subfigure}
\caption{Contours of the normalized vertical displacement $u_z/\ell_T(10^{-2})$ the sheet-pile wall in homogeneous domains (inverse analysis).}
\label{fig:Stress_inversion1}
\end{figure}

\begin{figure}[!t]
\centering
\begin{subfigure}{0.48\textwidth}
    \centering
    \includegraphics[trim={0 0 0 0},width=0.99\linewidth]{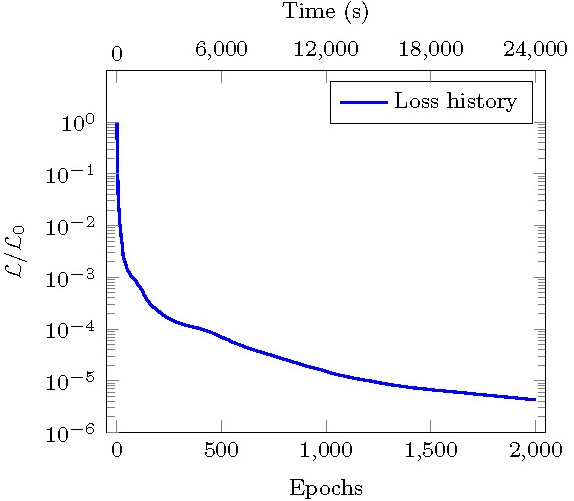}
\end{subfigure}
\caption{Training data history for the pile in layered soil.}
\label{fig:loss-inversion-layered}
\end{figure}

\begin{figure}[!b]
\centering
\begin{subfigure}{0.3\textwidth}
    \centering
    \includegraphics[trim={0 0 0 0},width=0.95\linewidth]{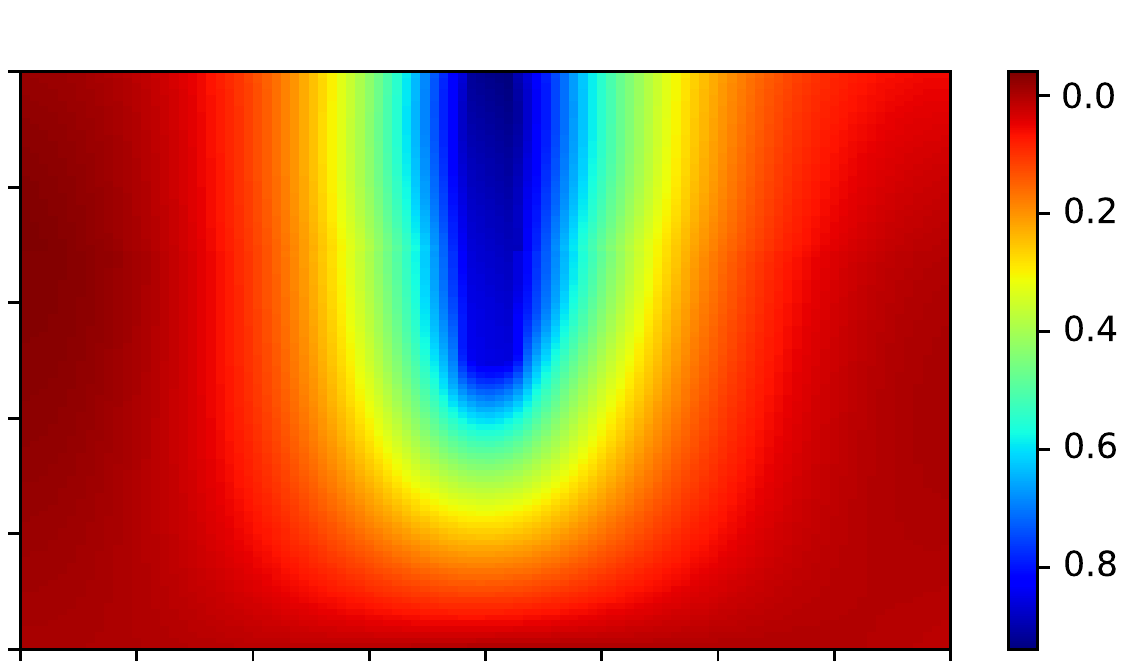}
    \caption{PINN solution}
    \label{fig:EX-b}
\end{subfigure}
\begin{subfigure}{0.3\textwidth}
    \centering
    \includegraphics[trim={0 0 0 0},width=0.92\linewidth]{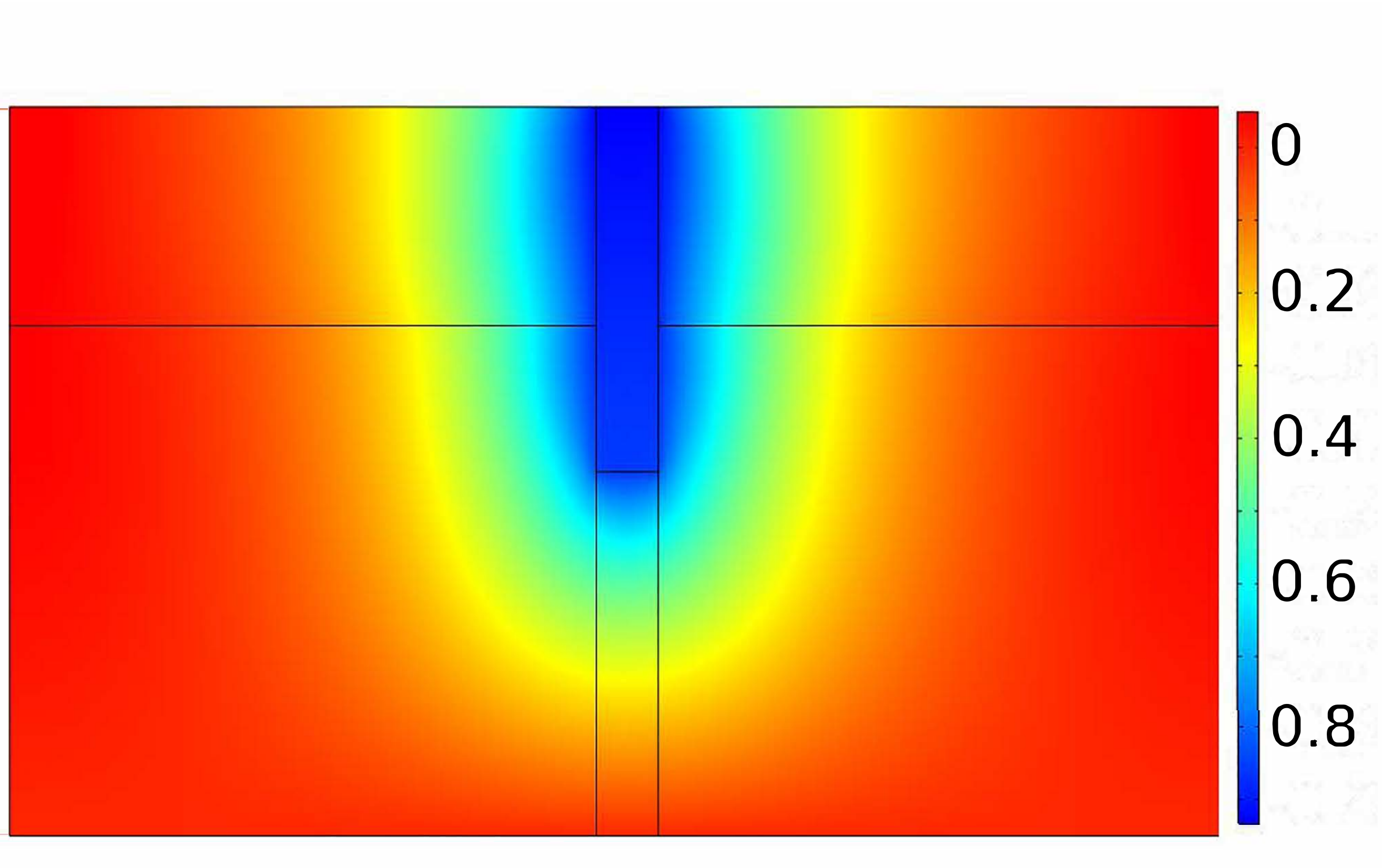}
    \label{fig:EX4-a}
    \caption{Reference solution}
\end{subfigure}
\caption{Contours of the normalized vertical displacement $u_z/\ell_T (10^{-2})$ for the pile in layered soil.}
\label{fig:stress-inversion-layered}
\end{figure}

In this example, we evaluate the performance of PINNs for the identification of model parameters involved in pile-soil systems. For this sake, the interaction of a single pile with the surrounding homogeneous soil medium under the plane strain condition is considered first. The material properties considered in here are: Young's Modulus of Elasticity, $E_P=5 \text{ GPa}$, $E_S=0.5, 0.1, 0.02 \text{ GPa}$; Poisson's ratio, $\nu_S=\nu_P=0.25$, stiffness ratios $\eta=E_P/E_S=10,\ 50,\ 250$. The input data-set is due to the longitudinal strain (or equivalently stress) profile along the pile, which is obtained by means of optical fiber strain-sensing technique in practice (e.g., see \cite{mohamad2011performance,mohamad2012monitoring}). In lieu of field data, a synthetic data-set is generated through the high-fidelity FEM solution of the same problem using COMSOL software. To emulate the limitations encountered in practice, merely a 1D profile along the center line of the pile is extracted from the FEM solution. This input data, in turn, is employed for the inversion of the Young's Modulus of the surrounding soil (i.e., $E_S$). Sampling grid involves 3000 sampling points over each solution domain $\Omega_{P}$ and $\Omega_{S}$, in conjunction with an input data consisting of 2000 points (8000 points in total).  It is noteworthy that a similar study is not feasible by the use of conventional deep learning considering the sparsity of the input data in this problem. Here, we demonstrate the versatility of PINNs in handling such study with extremely limited data-set.

The same governing equations presented in the previous example are applied in here for the inverse analysis of piles in homogeneous formations (i.e., Eqs. \eqref{GovEqCarEx2}-\eqref{eq:contact2}). However, the loss function is now updated by the inclusion of an extra term corresponding to the available input data as

\begin{equation}\label{eq:loss_sum21}
\begin{split}
&\mc{L}_T = \mathcal{L}_\Omega + \mathcal{L}_{\Gamma_\text{B.C.}} +\mathcal{L}_{\Gamma_\text{Cont}}+\mathcal{L}_{\Omega_{\sigma_{zz}^P}}\,\,\,,\\
&\mathcal{L}_{\Omega_{\sigma_{zz}^P}} = \lambda_{11} \left\|\ \sigma_{zz}^P-{\Bar{\sigma}^P_{zz}}\right\|_{\text{on } {\Omega_P}}\,\,\,,
\end{split}
\end{equation}
in which $\Bar{\sigma}_z^P$ is the input data-set. 

In Fig. \ref{fig:loss_inversion1}, the training history of the normalized loss versus epochs and time is presented for the identification study. As can be seen, the relative error norm has reached below $10^{-4}$ within $1000$ epochs. This indicates a substantial reduction in convergence rate in comparison to the forward study, which is expected in inverse solutions. In Fig. \ref{fig:Stress_inversion1}, contours of the normalized vertical displacement field $u_{z}/\ell_T$ are presented for the backward solution and compared to the reference FEM solution using COMSOL. Evidently, the PINNs results are in excellent agreement with the reference solution. In table \ref{table: discovery-rectangular} the exact values of the Young's modulus of the soil inverted in each case are reported. Notably, the deduced values in all cases lie within $5\%$ variation from the exact values. 

In the last example, the application of PINNs to identify the material parameters of layered soils based on the inverse analysis of the axial strain profile measured/obtained along the pile is demonstrated. Consider a two-layered soil with the thickness ratios of $\ell_1 / \ell_t=0.25$ and $\ell_2/\ell_t=0.75$. The material properties of the domain are: Young's Moduli of Elasticity, $E_P=5 \text{ GPa}$, $E_{S_1}=0.1 \text{ GPa}$, $E_{S_2}=0.02 \text{ GPa}$; Poisson's ratio, $\nu_P=\nu_{S_1}=\nu_{S_2}=0.25$. Here, the equations governing the response of homogeneous soils (i.e., Eq.s \eqref{GovEqCarEx2}-\eqref{eq:contact2}) need to be extended by inclusion of both soils layers (i.e., $\alpha=S_1, S_2$). In particular, the contact constraints (i.e., Eq. \eqref{eq:contact2}) need to be imposed across the interface of the pile with each soil layer as well as between the soil layers itself (see Eq. \eqref{eq:contact}). Such derivations are straightforward task, which is not presented here for the sake of brevity. 

PINNs are employed to extract the Young's moduli of both soils (i.e., $E_{S_1}$ and $E_{S_2}$). This inversion is again carried out by using a 1D data-set involving the strain (stress) profile across the center line of the pile. Six neural networks are elaborated for the PINNs solution in this example as

\begin{equation}\label{eq:NN7}
\begin{split}
u_x^P &\simeq \mathcal{N}_{u_x}^P(x,z)\,\,\,, \ \ \ \  u_z^P\simeq \mathcal{N}_{u_z}^P(x,z)\,\,\,, \\
u_x^{S_1} &\simeq \mathcal{N}_{u_x}^{S_1}(x,z)\,\,\,, 
\ \ \ u_z^{S_1}\simeq \mathcal{N}_{u_z}^{S_1}(x,z)\,\,\,, \\
u_x^{S_2} &\simeq \mathcal{N}_{u_x}^{S_2}(x,z)\,\,\,, 
\ \ \ u_z^{S_2} \simeq \mathcal{N}_{u_z}^{S_2}(x,z)\,\,\,, \\
\end{split}
\end{equation}
which consist of the same architecture as the previous case. The loss function is extended to incorporate both layers of the soil in conjunction with the additional contact constraints as

\begin{equation}
\begin{split}
\label{eq:loss_sum3}
&\mc{L}_T = \mathcal{L}_\Omega + \mathcal{L}_{\Gamma_\text{B.C.}} +\mathcal{L}_{\Gamma_\text{Cont}}+\mathcal{L}_{\Omega_{\sigma_{zz}^P}}\,\,\,,
\\
\\&\mathcal{L}_\Omega = \lambda_1 \left\|\mc{P}_{xx}^P \textbf{u}^P\right\|_{\text{on }\Omega_P}+ \lambda_2 \left\|\mc{P}_{xx}^{S_1}\textbf{u}^{S_1}\right\|_{\text{on }\Omega_{S_1}}+ \lambda_3 \left\|\mc{P}_{xx}^{S_2}\textbf{u}^{S_2}\right\|_{\text{on }\Omega_{S_2}}
\\
&\ \ \ \ + \lambda_4 \left\|\mc{P}_{zz}^P \textbf{u}^P\right\|_{\text{on }\Omega_P}+\lambda_5 \left\|\mc{P}_{zz}^{S_1} \textbf{u}^{S_1}\right\|_{\text{on }\Omega_{S_1}}+\lambda_6 \left\|\mc{P}_{zz}^{S_2} \textbf{u}^{S_2}\right\|_{\text{on }\Omega_{S_2}}\,\,\,,
\\
\\
&\mathcal{L}_{\Gamma_\text{B.C.}} =\lambda_7 \left\| \mc{B}_{xx}^P \textbf{u}^P - g_{xx}^P \right\|_{\text{on } \Gamma_P \setminus (\Gamma_{S_1} \cup \Gamma_{S_2})} + \lambda_8 \left\| \mc{B}_{xx}^{S_1} \textbf{u}^{S_1} - g_{xx}^{S_1} \right\|_{\text{on} \Gamma_{S_1} \setminus (\Gamma_{P} \cup \Gamma_{S_2})}
\\
&\ \ \ \ \ \ \ \ +\lambda_9 \left\| \mc{B}_{xx}^{S_2} \textbf{u}^{S_2} - g_{xx}^{S_2} \right\|_{\text{on} \Gamma_{S_2} \setminus (\Gamma_{P} \cup \Gamma_{S_1})}
+\lambda_{10} \left\| \mc{B}_{zz}^P \textbf{u}^P - g_{zz}^P \right\|_{\text{on } \Gamma_P \setminus (\Gamma_{S_1} \cup \Gamma_{S_2})}
\\
&\ \ \ \ \ \ \ \ +\lambda_{11} \left\| \mc{B}_{zz}^{S_1}  \textbf{u}^{S_1}  - g_{zz}^{S_1}  \right\|_{\text{on } \Gamma_{S_1} \setminus (\Gamma_{P} \cup \Gamma_{S_2}) }+\lambda_{12} \left\| \mc{B}_{zz}^{S_2}  \textbf{u}^{S_2}  - g_{zz}^{S_2}  \right\|_{\text{on } \Gamma_{S_2} \setminus (\Gamma_{P} \cup \Gamma_{S_1})}\,\,\,,
\\
\\
&\mathcal{L}_{\Gamma_\text{Cont}} =\lambda_{13} \left\| \textbf{u}^P - \textbf{u}^{S_1} \right\|_{\text{on } _{\Gamma_P \cap \Gamma_{S_1}}}+\lambda_{14} \left\| \textbf{u}^P - \textbf{u}^{S_2} \right\|_{\text{on } _{\Gamma_P \cap \Gamma_{S_2}}}+\lambda_{15} \left\| \textbf{u}^{S_1} - \textbf{u}^{S_2} \right\|_{\text{on } _{\Gamma_{S_1} \cap \Gamma_{S_2}}}
\\
&\ \ \ \ \ \ \ \ + \lambda_{16} \left\| \textbf{t}^P - \textbf{t}^{S_1} \right\|_{\text{on } _{\Gamma_P \cap \Gamma_{S_1}}}+\lambda_{17} \left\| \textbf{t}^P - \textbf{t}^{S_2} \right\|_{\text{on } _{\Gamma_P \cap \Gamma_{S_2}}}+\lambda_{18} \left\| \textbf{t}^{S_1} - \textbf{t}^{S_2} \right\|_{\text{on } _{\Gamma_{S_1} \cap \Gamma_{S_2}}}\,\,\,,\\
\\
&\mathcal{L}_{\Omega_{\sigma_{zz}^P}} = \lambda_{19} \left\| \sigma_{zz}^P-\Bar{\sigma}_{zz}^P\right\|_{\text{on } {\Omega_P}}\,\,\,,
\end{split}
\end{equation}

Sampling grid involves 3000 sampling points over $\Omega_P$, 2000 points in each of the soil layers $\Omega_{S_1}$ and $\Omega_{S_2}$, and 2000 points due to the input data-set, which is summed at 9000 points in total. In Fig. \ref{fig:loss-inversion-layered}, the training history of the normalized loss is presented. As can be seen, the convergence rate versus epochs has been relatively fast. Still, given the number of losses has expanded dramatically, the training is conducted significantly slower with respect to the case of inversion in homogeneous soils. Contours of the normalized vertical displacement field $u_{z}/\ell_T$ are depicted in Fig. \ref{fig:stress-inversion-layered} for both PINNs and reference solution by COMSOL. Evidently, excellent agreement is observed between the PINNs results and FE simulation. Finally, in Table \ref{table: layered} the inverted Young's modulus for each soil layer is reported. The inverted values are in very good agreement with the precise amounts. This further demonstrates the excellent performance of PINNs in complex parametric studies pertaining to extremely limited input data.
\begin{table}[!b]
\footnotesize
\centering
\caption{Inversion of soil Young's Modulus for the pile embedded in homogeneous domains (unit: GPa).}
\begin{tabular}{cccc}
\Xhline{2\arrayrulewidth} Analysis &
  {$E_P$} & {$E_{S}^{\text{pre}}$} & {$E_{S}^\text{exact}$} \\ 
\Xhline{2\arrayrulewidth}
 {Analysis 1.} & 5.0 & 0.0213 & 0.02\\
 \hline 
 {Analysis 2.} & 5.0 & 0.105 & 0.10\\
 \hline
 {Analysis 3.} & 5.0  & 0.525 & 0.5 \\
\Xhline{2\arrayrulewidth}
\end{tabular}
\label{table: discovery-rectangular}
\end{table}

\begin{table}[!b]
\footnotesize
\centering
\caption{Inversion of soil Young's Modulus for the pile embedded in layered formation (unit: GPa).}
\begin{tabular}{cccccc}
\Xhline{2\arrayrulewidth} Analysis &
  {$E_P$} & {$E_{S_1}^ \text{pre}$} & {$E_{S_1}^ \text{excat}$}& {$E_{S_2}^\text{pre}$} & {$E_{S_2}^ \text{excat}$} \\ 
\Xhline{2\arrayrulewidth}
 {} & 5.0 & 0.095 & 0.1 & 0.023 & 0.02\\
 \Xhline{2\arrayrulewidth}
\end{tabular}
\label{table: layered}
\end{table}

\section{Conclusions}
\label{S:5 (Conclusions)}
A physics-informed deep learning framework is presented for the analysis of pile-soil interaction under vertical loading. In the framework, a domain-decomposition multi-network model is introduced to deal with the sharp discontinuities in the strain field at the interfaces of pile-soil regions and soil layers. The framework is trained by minimizing the loss function defined in terms of the equilibrium equations and the boundary conditions governing the pile-soil interaction problem. Several examples are provided to demonstrate the performance of the framework in the analysis of single piles embedded in homogeneous and layered soils under axisymmetric and plane strain conditions. Essential features of the framework are validated by comparing the PINN results with the results obtained from the finite element analysis. Good agreement is observed between the PINN and FEM results in all the cases considered. The application of the model for the inverse analysis and parameter identification of pile-soil interaction is also presented. In the examples provided, the localized data acquired along the pile length–possibly obtained via fiber optic strain sensing– is used for the inversion of soil parameters in both homogeneous and layered formations. As expected, a substantial reduction in the convergence rate is observed in the inverse analysis in comparison to the forward study. However, 
it is seen that the proposed PINN framework is able to identify the material parameters quite efficiently.
\bibliographystyle{model1-num-names.bst}
\bibliography{References.bib}
\end{document}